\documentclass[twocolumn]{aastex631}
\usepackage{newtxtext,newtxmath}

\usepackage{float}
\usepackage{makecell}
\usepackage{threeparttable}
\usepackage{hyperref}
\usepackage{cleveref}
\usepackage{ulem}
\crefname{table}{Tab.}{Tables}
\crefname{figure}{Fig.}{Figures}

\usepackage{graphicx}

\submitjournal{AJ}

\shorttitle{Revision of the Detached Eclipsing System IR Cas}
\shortauthors{Kamenec et al.}

\begin{document}

\title{Revision of the Detached Eclipsing System IR\,Cas from TESS Observations, Ground-Based Photometry and Spectroscopy}                      

\author[0009-0002-2976-0719]{Mat\'u\v{s} Kamenec}
\affiliation{Institute of Physics, Faculty of Science, Pavol Jozef \v{S}af\'arik University, 04001 Ko\v{s}ice, Slovakia}
\email{matus.kamenec@student.upjs.sk}

\author[0000-0003-1478-3256]{Pavol Gajdo\v{s}}
\affiliation{Institute of Physics, Faculty of Science, Pavol Jozef \v{S}af\'arik University, 04001 Ko\v{s}ice, Slovakia}

\author[0000-0002-2798-6944]{Martin Va\v{n}ko}
\affiliation{Astronomical Institute of the Slovak Academy of Sciences, Tatransk\'a Lomnica, SK-059~60 Vysok\'e Tatry, Slovakia}

\author[0009-0006-6663-4200]{Reddy Charan Reddy Munagala}
\affiliation{Astronomical Institute of the Slovak Academy of Sciences, Tatransk\'a Lomnica, SK-059~60 Vysok\'e Tatry, Slovakia}

\author[0000-0002-7602-0046]{Marek Skarka}
\affiliation{Astronomical Institute, Czech Academy of Sciences, Fri\v{c}ova 298, 25165, Ond\v{r}ejov, Czech Republic}

\author[0000-0002-7204-9220]{{{\v{S}}tefan} Parimucha}
\affiliation{Institute of Physics, Faculty of Science, Pavol Jozef \v{S}af\'arik University, 04001 Ko\v{s}ice, Slovakia}

\begin{abstract}
We present a new photometric and spectroscopic analysis of detached eclipsing binary IR Cas based on TESS observations, supplementary ground-based photometry in Sloan $g^\prime$, $r^\prime$, and $i^\prime$ filters, and newly obtained radial velocity measurements. The updated orbital and physical parameters of the system were derived using combined light-curve and radial-velocity modeling. The resulting solution indicates that both components are main-sequence stars with masses of approximately $1.32$\,M$_{\odot}$ and $1.05$\,M$_{\odot}$. We investigated in detail the fact, that the TESS light curves exhibit asymmetries near the maxima, which were reproduced by introducing a cool spot that moves on the surface of the secondary component. Long-term analysis of times of minima revealed quasi-periodic variations in the O$-$C diagram that can be interpreted as a light-time effect due to a possible third body with an orbital period of about 38 years. The positions of both components in the mass--radius diagram agree well with empirical relations for detached main-sequence binaries and do not indicate substantial deviations from standard stellar evolution. Overall, IR Cas appears to be an evolutionarily representative detached eclipsing system with moderate indications of stellar activity.
\end{abstract}


\section{Introduction}

For eclipsing binary stars (EB), the changes in their brightness are caused by mutual eclipses, which can be observed when the observer’s line of sight lies close to the plane of their orbit. Eclipsing binaries constitute a key class of variable stars, providing a unique opportunity to determine the physical parameters of their components by analyzing their light curves (LC) and variations in radial velocities (RV) \citep{Hilditch_2001}. Accurate knowledge of these parameters—such as mass, radius, and effective temperature—is crucial for theoretical models of stellar evolution, statistical studies, and broader aspects of stellar astrophysics. Moreover, additional phenomena, including pulsations, star spots, and variations in eclipse timings, are frequently observed in the light curves of EBs.

Thanks to ongoing ground-based observations and satellites such as Gaia \citep{Gaia2016}, Kepler \citep{2010Sci...327..977B}, and TESS \citep{2015JATIS...1a4003R}, the amount of available photometric data has increased significantly. The International Variable Star Index (VSX) now lists nearly one million EBs \citep{2006SASS...25...47W}. About 3000 EBs in the Kepler field have been well studied using data from this mission \citep{Kirk_2016}. Additionally, more than 10000 EBs have been identified by the TESS mission \citep{Kostov_2025}, while the Gaia mission has detected almost 2.1 million EBs \citep{Mowlavi_2023}. High-precision photometry from Kepler, Kepler-K2, and TESS enables the study of the finest effects in these systems over long, uninterrupted timescales. However, many eclipsing binaries still lack detailed analyzes based on modern multicolor photometry and spectroscopic observations. IR Cas belongs to this group of relatively poorly studied systems, for which only limited modern  observational material is currently available. Additional photometric and
spectroscopic observations may therefore contribute to refining the system parameters and improving our understanding of its orbital and physical properties. For this reason, IR Cas was selected as part of a long-term monitoring program of eclipsing binaries accessible with modest observational equipment.

According to previous studies, IR Cas belongs to the group of close EBs, in which the components nearly fill their Roche lobes. From a phenomenological point of view, their light curves are of the $\beta$\,Lyrae type and may exhibit secondary effects. The first photometric analysis was published in 2014 \citep{Li_2014}. According to the SIMBAD database \citep{2000A&AS..143....9W} IR Cas is classified as F4, and is a short-period eclipsing binary (EB) with an orbital period of only about 16 hours (0.680685 days). The first mention of this object can be found in \citet{1943AN....274...36H}. The V magnitude is 11.14, and the parallax measured by Gaia is 2.83 mas, yielding a distance of 353 pc. The Gaia DR3 catalog \citep{2023A&A...674A...1G} does not provide information on the system’s temperature, but Gaia DR2 \citep{2018A&A...616A...1G} estimated it to be 5780~K, which does not correspond to the reported spectral type. Additional parameters are given in \cref{tab:ircas_summary}.

\begin{table}
\centering
\caption{Identifications and basic parameters of IR~Cas.}
\label{tab:ircas_summary}
\begin{tabular}{ll}
\hline
\textbf{Parameter} & \textbf{Value} \\
\hline
Identifiers & Gaia~DR3~1996353476965736448 \\
& 2MASS~J23065237+5404521 \\
& GSC~03998-02007 \\
& TIC~314823520 \\
R.A. (ICRS, J2000) &
23:06:52.374 \\

Dec (ICRS, J2000) &
+54:04:52.230 \\


Parallax (mas) &
$2.8371 \pm 0.0125$ \\

Distance (pc) &
352.473$\pm$1.553 \\

$\mu_\alpha \cos\delta$ (mas\,yr$^{-1}$) &
$17.262 \pm 0.013$ \\

$\mu_\delta$ (mas\,yr$^{-1}$) &
$-5.968 \pm 0.011$ \\


Spectral type &
F4 \\


$B$ (mag) &
$11.75 \pm 0.09$ \\

$V$ (mag) &
$11.14 \pm 0.08$ \\

$G$ (mag) &
$10.885 \pm 0.012$ \\

$J$ (mag) &
$9.515 \pm 0.021$ \\

$H$ (mag) &
$ 9.251 \pm 0.015$ \\

$K$ (mag) &
$ 9.181 \pm 0.016$\\



\hline
\end{tabular}

\vspace{0.4ex}
\footnotesize
\textit{Notes.}
All parameters were compiled from the SIMBAD database, aggregating data from
Gaia~DR3 \citep{2023A&A...674A...1G}, 2MASS \citep{2006AJ....131.1163S}, AllWISE \citep{2010AJ....140.1868W}, IRAS \citep{1988iras....1.....B} and related catalogues.
\end{table}


In recent years, two in-depth photometric analyzes have been conducted. \citet{Li_2014} performed \textit{BVRI} photometry and obtained a mass ratio of \textit{q} = 0.851, adopting $T_1$ = 6750~K and obtaining the corresponding system parameters.
They also investigated period variations in the O–C diagram and, assuming a light-time effect, determined that the period is approximately 39.7~yrs and the amplitude is approximately 0.0153~days. Based on their observations, they concluded that IR Cas is a semi-detached system in which the primary fills its Roche lobe and can be classified as a near-contact binary. The absolute parameters of the system were derived under the assumption that the primary component is a standard main-sequence star with $M_1$ = 1.43 $M_\odot$.
Based on the continuous decrease in the orbital period, they concluded that mass could be transferred from the primary to the secondary. For the suspected third body, the estimated values are $M_3$ = 0.49 $M_\odot$ and $a_3$ = 17.09 au. They assumed the third body to be a main-sequence star, and its estimated mass suggests a spectral type around M0V. In such a case, the tertiary component would contribute less than 1 per cent to the system's total light, making its photometric and spectroscopic detection with ground-based telescopes extremely difficult. They also found that the system will evolve from its current semi-detached configuration into a contact phase. Assuming a mean orbital period of 0.4 days for contact binaries, the system is expected to become a contact binary after approximately \(1.2 \times 10^{6}\) years.


\citet{NELSON2022101770} analyzed the period variations and reproduced the \textit{O–C} diagram using a light-time effect model. Based on low-resolution spectroscopy, he derived the radial velocities and obtained the basic orbital parameters and mass ratio.
In 2017 and 2018, \citet{NELSON2022101770} conducted photometric observations and performed a photometric analysis of these data. 
Comparison with the results of the photometric analysis of \citet{Li_2014} reveals remarkable agreement; however, they concluded that this is a semi-detached system, while \citet{NELSON2022101770} concluded that it is a detached system with both stars evolved. They regard their values as preferable because they were derived from RV data. In connection with the LiTE effect, they performed a new fit, this time omitting the early photographic observations. From the wide discrepancy among the derived parameters, they concluded that new measurements of eclipse timings collected over many additional decades will be required to resolve this issue and that the current conclusions regarding period changes are premature.

We aimed to utilise data currently available from ground-based observations, from the TESS satellite, and from our own observations carried out with the instrumentation at the Kolonica Saddle Observatory. The O-C diagram constructed from TESS data allowed us to derive an improved ephemeris. We also constructed a long-term O-C diagram to investigate the suspected presence of a third body, performed spectroscopic observations to determine radial velocities, and derived absolute parameters of the system. Moreover, we constructed a spot model and analyzed its behavior on the component surface. Finally, we constructed several relations between physical parameters using a broader sample of similar systems and discussed the selected evolutionary characteristics of the system studied.

\section{TESS data acquisition and processing}
\label{tess}




The TESS satellite provides full-frame image (FFI) observations of IR Cas in six sectors, of which three (sectors 16, 17, and 57) were found to be usable. In the remaining sectors, the data was unusable, apparently due to instrumental issues. To obtain photometric data, we used custom standalone scripts based on the lightkurve package \citep{2018ascl.soft12013L}, as well as their online version with a web interface \footnote{\href{https://astronomy.science.upjs.sk/igebc/tess}{https://astronomy.science.upjs.sk/igebc/tess}}. The aperture in the FFI cuts was set manually. 


The flux was normalized to its maximum value, and then the data was detrended, primarily by subtracting the background. In this way, we obtained three light curves from three sectors, suitable for further analysis. 

\begin{figure}
	\centering
      \includegraphics[width=1.0\columnwidth]{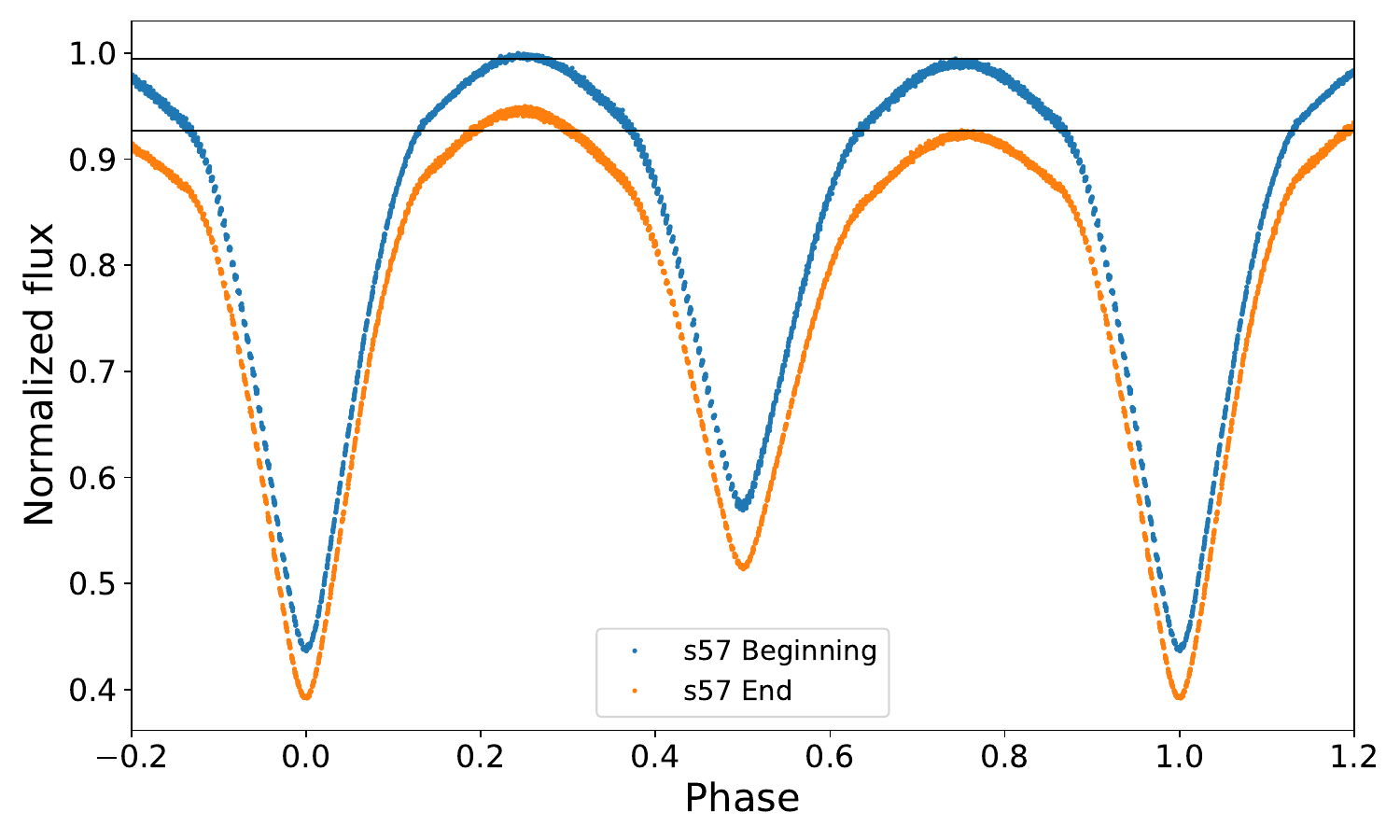}
	\caption{Phase curves of IR Cas, beginning of sector 57 (blue) and end of sector 57 (orange). A flux shift of 0.05 was applied between
    the phase curves.}
	\label{FIG:phasecurves}
\end{figure}

Based on the shape of the light curve, this EB can be classified as a $\beta$ Lyrae type, which is consistent with the available information on the star \citep[e.g. VSX,][]{1978IBVS.1446....1K}. In the TESS light curves, we observe significant changes in the maxima, which may be influenced by stellar spots \citep[e.g. ][]{Gajdos_2024}. However, within a single sector, we can also observe variations in the differences between the maxima. 
Therefore, in Sector 57, we constructed two phase curves corresponding to the beginning and end of the light curve, each covering approximately 10 orbital periods, in which the differences in the maxima are clearly visible (\cref{FIG:phasecurves}). These variations can also be observed in sectors 16 and 17, despite the smaller number of data points caused by the longer exposure time.

\section{Ground-based photometry}
\label{phot}

Our observations were carried out with a Newton-type 200/1200 telescope on a Sky Watcher NEQ-6 Pro equatorial mount, equipped with a Moravian Instruments G2-8300 CCD camera with Sloan/SDSS $g'$, $r'$, $i'$ filters. The telescope is located at the Kolonica Saddle Observatory. However, it can be operated fully remotely. 

Multi-channel photometry was performed over three consecutive nights, from 5 to 7 November 2025  (\cref{tab:ground_obs}). We used an exposure time of 30 seconds in all filters. The acquired images were processed and calibrated using the Muniwin software package \citep{2014ascl.soft02006H}. For the purposes of differential photometry, the comparison star TYC 3998-2101-1 (GSC 03998-02101) with a color index $B-V = 0.08$ and $V$ magnitude of 10.46 was selected \citep{2000A&A...355L..27H}. No explicit color-term correction was applied to the differential photometry. Although the comparison star is significantly bluer than the IR Cas, the light curve analysis presented in this work is based primarily on relative brightness variations and the shape of the light curves rather than on absolute calibrated magnitudes. Therefore, any residual color-dependent effects are expected to have only a minor influence on the derived geometric parameters of the system.
From the three obtained light curves, phase curves were constructed for each filter (see Fig.~\ref{fig:lc-fit}).
The differential photometry obtained is available as supplementary material.

\begin{table}
\centering
\caption{The observational log of ground-based photometric observations during individual nights.}
\label{tab:ground_obs}
\footnotesize
\begin{tabular}{lccc}
\hline
Date & Time (UT) & Phase & No. Obs. ($g'/r'/i'$) \\
\hline
2025 Nov 05 & 16:08--02:59 & 0.54--0.05 & 251 / 320 / 320 \\
2025 Nov 06 & 16:24--02:59 & 0.87--0.52 & 304 / 301 / 299 \\
2025 Nov 07 & 16:57--00:41 & 0.38--0.85 & 182 / 181 / 176 \\
\hline
\end{tabular}
\footnotesize{Note. Only observations used in the following analysis are listed.}

\end{table}

\section{O-C diagram}
\label{oc}

First, all available TESS observations were used to derive a precise linear ephemeris. Individual primary and secondary minima were identified, and their mid-times were determined by approximating their profiles with a Gaussian function and fitting them using the Monte Carlo method. The resulting times (available as supplementary material to this manuscript) were analyzed with our \texttt{OCFit} software \citep{ocfit}, giving the linear ephemeris for the primary minima in the following form:

\begin{equation}
\label{eq:eph}
T_{\rm I} = {\rm BJD}~2\,459\,853.605433 (37) + 0.680687560 (72) \times E\,,
\end{equation}
where $E$ is the epoch of the observation, and uncertainties are given in parentheses.

\begin{figure}
	\centering
	\includegraphics[width=1.0\columnwidth]{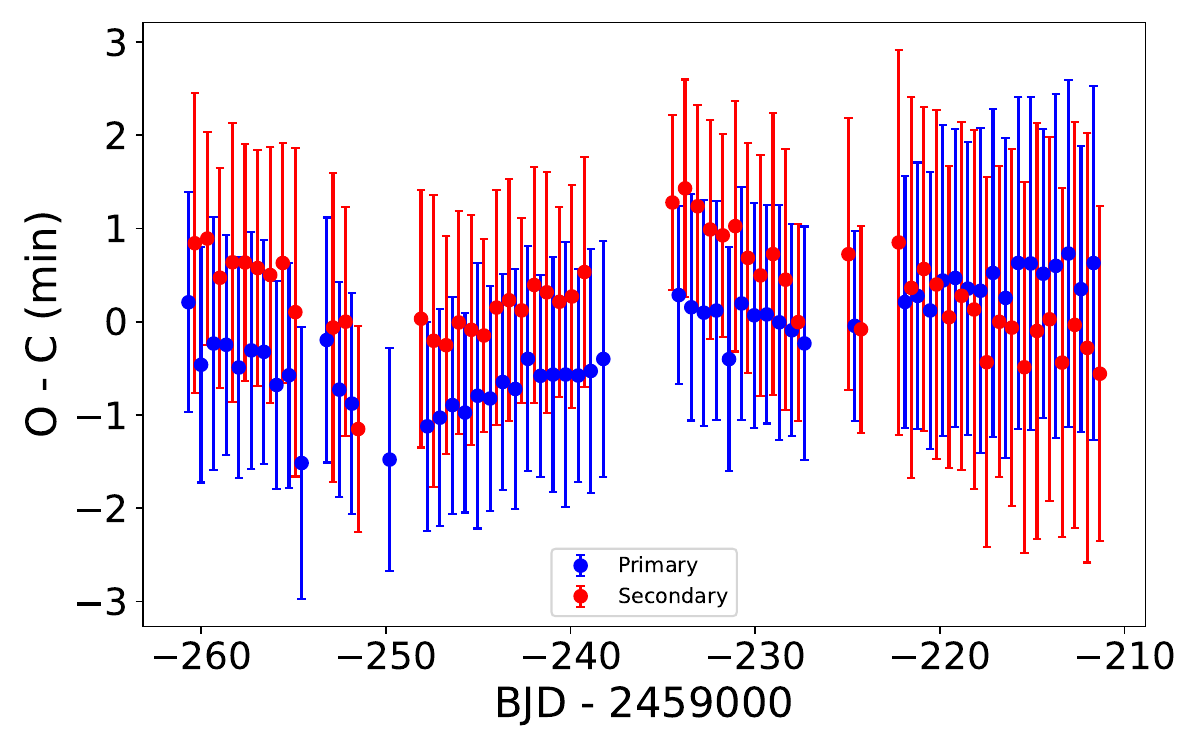}
    \includegraphics[width=1.0\columnwidth]{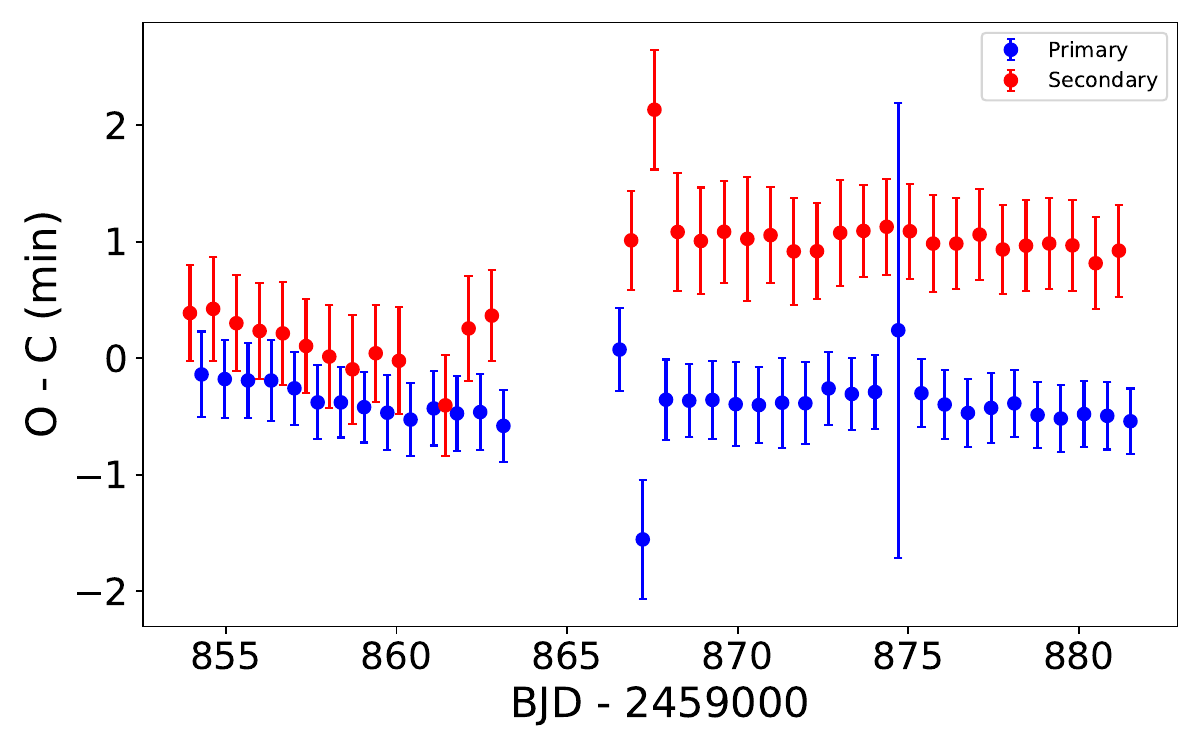}
	\caption{O-C diagram based on TESS data, sectors 16 and 17 at the top, sector 57 at the bottom.}
	\label{fig:oc-tess}
\end{figure}

Using improved ephemeris, we constructed the O-C diagram shown in Fig.~\ref{fig:oc-tess}. The O-C amplitudes of the primary and secondary minima are on the order of 1 minute. The times of the primary and secondary minima exhibit quasi-periodic or chaotic, mostly anti-phased, short-term variations that may be attributed to the presence of a spot \citep[e.g.][]{Gajdos_2024}.

\begin{figure}
	\centering
	\includegraphics[width=1\columnwidth]{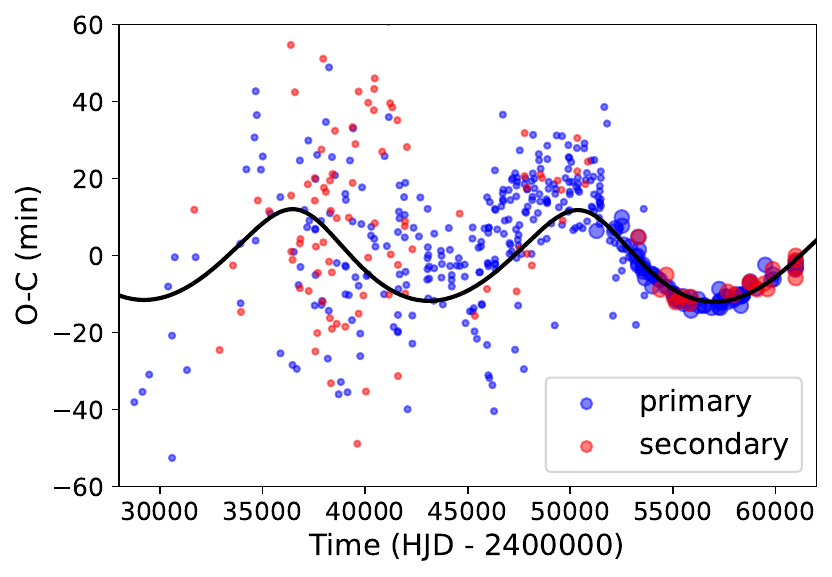}
	\caption{O-C diagram based on available observations. The size of the points represents the weights of the data points according to the observation method used.}
	\label{fig:oc}
\end{figure}

Together with historical data (over a century), we have collected nearly 500 minima, including photographic, photoelectric, visual, and CCD observations available at the \mbox{O-C} gateway\footnote{https://var.astro.cz/} of the Czech astronomical society \citep{ocgate}. We excluded 2 photographic measurements from 1906 because they were significantly out of trend. We weight observations on the basis of the different observing methods used. We also added our ground-based times of the minima (Tab.~\ref{tab:times}). For TESS observations, we selected one primary and one secondary minimum per sector, with an average O-C value corresponding to the selected sector. This approach extends the time baseline of historical data without adding unwanted scatter, which could otherwise be caused by the short-term chaotic behavior of minimum times in the TESS observations.

\begin{table}[h]
    \centering
    \caption{Determined minima times used in analysis of long-term period changes for primary (P) and secondary minima (S). Source data are TESS observations (one minima of each time per sector) and ground-based photometry.}
    \label{tab:times}
    \footnotesize
    \begin{tabular}{ccc}
    \hline
    Minima time (HJD) & Type & Source \\
    \hline
    2458746.807324 & P & TESS - S16 \\ 
    2458751.912639 & S & TESS - S16 \\  
    2458778.119235 & P & TESS - S17 \\ 
    2458775.736625 & S & TESS - S17 \\ 
    2459867.899625 & P & TESS - S57 \\ 
    2459868.921603 & S & TESS - S57 \\ 
    2460969.251290 & P & Ground \\
    2460984.565212 & P & Ground \\
    2460985.588104 & P & Ground \\ 
    2460986.269674 & P & Ground \\
    2460986.610517 & S & Ground \\
    2460987.291993 & S & Ground \\
    2460998.521283 & P & Ground \\
    \hline
    \end{tabular}    
\end{table}

We determined the linear ephemeris on this long time scale to be
\begin{equation}
\label{eq:eph-all}
T_{\rm I} = {\rm HJD}~2\,459\,853.60846 (135)  + 0.680686518 (41) \times E\,.
\end{equation}
In the O-C diagram, the periodic variation is clearly visible (see Fig.~\ref{fig:oc}). Its amplitude is around 10 minutes over a period of about 40 years. These long-term changes were already observed by \cite{Li_2014} and \cite{NELSON2022101770}. Moreover, \cite{Li_2014} considered a period decrease at the level of about $-1.28 \cdot 10^{-7}$~d/yr. However, this trend is based only on two observations from the beginning of the 20$^{\rm th}$ century.

\begin{table*}
\caption{Parameters of the 3$^{\rm rd}$ body determined in O-C analysis -- orbital period $P_3$, time of pericenter passage $t_{03}$, amplitude on O-C diagram $A$, projected semi-major axis of EB around baricenter $a_{12}\sin i_3$, eccentricity $e_3$, argument of pericenter $\omega_3$, mass function $f(M_3)$, mass $M_3$ and projected mass $M_3\sin i_3$, distance between the 3$^{\rm rd}$ body and EB $a$. }
\label{tab:oc}
\centering
\begin{tabular}{lccc}
	\hline
	Parameter                  &       This work       &      \cite{Li_2014}     & \cite{NELSON2022101770} \\ \hline
	$P_3$ (yr)                 &   38.06 $\pm$ 0.72    &  39.7 $\pm$ 1.5   &          38.0           \\
	$t_{03}$ (HJD)             &  2450858 $\pm$ 1607   & 2448184 $\pm$ 317 &         2451230         \\
	$A$ (min)                  &  11.86  $\pm$  1.88   &    22 $\pm$ 19    &          11.52          \\
	$a_{12} \sin i_3$ (au)     &    1.46 $\pm$ 0.21    &        ---        &           ---           \\
	$e_3$                      &    0.25 $\pm$ 0.16    &  0.89 $\pm$ 0.22  &          0.36           \\
	$\omega_3$ (deg)           &   107.1 $\pm$ 41.6    &  10.5 $\pm$ 10.5  &          61.1           \\
	$f(M_3)$ (M$_\odot$)       & 0.00202 $\pm$ 0.00092 &        ---        &           ---           \\
	$M_3 \sin i_3$ (M$_\odot$) &   0.250 $\pm$ 0.040   &        ---        &           ---           \\
	$M_3^*$ (M$_\odot$)        &   0.251 $\pm$ 0.041   &        0.49       &           ---           \\
	$a^*$ (au)                 &    15.9 $\pm$ 3.4     &       17.09       &           ---           \\ \hline
\end{tabular}
\begin{flushleft}
$^*$ assuming coplanar orbit ($i = 85.05\deg$).
\end{flushleft}
\end{table*}

We used the standard Light-time effect model \citep{irwin-lite} in the \texttt{OCFit} code to analyze the long-term periodic changes on the O-C diagram. Our results (listed in Tab.~\ref{tab:oc}) are more consistent with the study of \cite{NELSON2022101770}, but we have a slightly longer time baseline. We can describe the observed periodic changes on the O-C diagram by the presence of an additional body with a mass of about 0.25~M$_\odot$ (assuming a coplanar orbit) with an orbital period of 38 years around the central EB. A similar mass corresponds to an M-type main-sequence star.

\section{Spectroscopy}
\label{spec}

We supplemented the photometric data with spectroscopic observations that provided further insight into the origin of the studied system. For this purpose, we used two spectrographs: the Ond\v{r}ejov Echelle Spectrograph (OES) attached to a 2-meter Perek telescope at the Ond\v{r}ejov observatory, and a clone of the \'echelle MUSICOS spectrograph (MUlti-SIte COntinuous Spectroscopy) mounted in the Nasmyth-Cassegrain focus of the 1.3m telescope at Skalnaté Pleso Observatory (SP).

\subsection{Instrumentation and Data Reduction}

The OES is a high-resolution instrument mounted on the 2-meter Perek telescope at the Ond\v{r}ejov Observatory, Czech Republic. It covers the wavelength range 3870--9200\,\AA~with a spectral resolution power R = 51600, centered at 5000\,\AA~\citep{oes1}. The long-term stability of the radial velocity is approximately 200--300~m/s \citep{oes2}. The limiting observation magnitude in the {\it V} filter is around 13~mag.

MUSICOS is a high-dispersion echelle spectrograph mounted on the 1.3-meter telescope located at Skalnat\'{e} Pleso Observatory, Slovakia (altitude of 1786\,m above sea level). The high altitude reduces atmospheric turbulence and reduces the water vapor content, improving the quality of the spectroscopic observations. MUSICOS covers a spectral range of 4250--7375\,\AA~with a spectral resolution of R = 25000--38500, depending on the focus \citep{musikos}. In addition, the authors report an RV stability of 100--200\,m/s.


The raw data from both spectrographs were reduced using a set of scripts based on the {\tt IRAF} and \texttt{pyraf} packages \citep{Tody1986,pyraf}. This pipeline includes dark-frame and flat-field corrections and removes cosmic hits using the code from \citet{Pych2004}. After extracting individual \'echelle orders, the wavelength calibration of the spectra is calculated using the ThAr lamp. The final step consists of continuum normalization and the combination of 2D spectra in 1D. For more details on the reduction process of MUSICOS spectra, see \cite{Pribulla2015, Garai2017}.

\subsection{Radial velocities}

The RVs were determined using the cross-correlation function (CCF) technique 
\citep[e.g.,][]{Griffin1967, Simkin1974, TonryDavis1979, Zverko2007}. We used the \texttt{iSpec} \citep{ispec2014,ispec2019} package for this purpose. We used synthetic spectra generated by the model of \citet{Coelho2014} as templates, with stellar parameters close to IR Cas. We selected only the spectral region around the magnesium triplet (5100–5200~\AA). We separately determined the CCF peaks corresponding to each system component identified in the spectra. If these two peaks overlap and the previous method is unusable, we fit a common CCF using a double-Gaussian function.

\begin{figure}
	\centering
	\includegraphics[width=1\columnwidth]{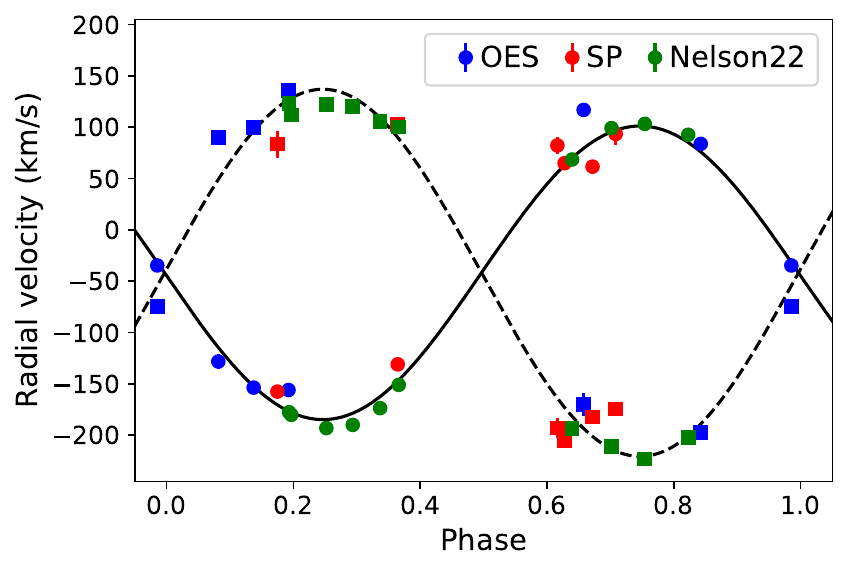}
	\caption{Radial velocities of both main components of IR Cas systems. Data are obtained by our measurements (Ond\v{r}ejov -- OES and Skalnat\'{e} Pleso observatory -- SP) and collected from the paper \citet{NELSON2022101770}. Full line represents theoretical curve for primary component and dashed for secondary one.}
	\label{fig:rv}
\end{figure}

\begin{table*}
\caption{Measured radial velocities of both components collected at the Ond\v{r}ejov observatory (OES) and the Skalnat\'{e} Pleso observatory (SP).}             
\label{tab:rv}      
\centering                        
\begin{tabular}{cccccc}
	\hline
 HJD - 2400000 & Phase & RV$_1$ (km/s)  &  RV$_2$ (km/s)  & S/N & Obs. \\ \hline
	 61028.21101  & 0.616 &  $82.2\pm7.9$  & $-193.2\pm9.7$  & 40  &  SP  \\
	 61029.27185  & 0.175 & $-157.5\pm4.6$ &  $83.4\pm13.0$  & 39  &  SP  \\
	 61030.26078  & 0.628 &  $64.9\pm6.8$  & $-205.1\pm5.0$  & 39  &  SP  \\
	 61031.44306  & 0.365 & $-131.0\pm1.9$ &  $102.8\pm3.8$  & 13  &  SP  \\
	 61032.35780  & 0.708 & $93.4\pm10.9$  & $-174.4\pm6.4$  & 22  &  SP  \\
	 61037.21383  & 0.842 &  $83.7\pm4.7$  & $-197.1\pm4.9$  & 43  & OES  \\
	 61052.28626  & 0.985 & $-34.7\pm3.3$  &  $-74.4\pm3.4$  & 15  & OES  \\
	 61059.19657  & 0.137 & $-153.7\pm4.7$ &  $99.7\pm7.1$   & 26  & OES  \\
	 61059.23402  & 0.192 & $-156.0\pm4.0$ &  $136.0\pm5.7$  & 36  & OES  \\
	 61060.23152  & 0.658 & $116.8\pm5.8$  & $-169.8\pm11.1$ & 24  & OES  \\
	 61058.19901  & 0.672 &  $61.5\pm5.1$  & $-182.4\pm5.5$  & 29  &  SP  \\
	 61061.20070  & 0.082 & $-128.3\pm7.2$ &  $89.7\pm5.6$   & 37  & OES  \\ \hline
\end{tabular}
\end{table*}

We collected 12 observations in total during December 2025 and January 2026 (listed in Tab.~\ref{tab:rv}). Uncertainties and scatter are large, mainly due to a short orbital period and faint brightness, resulting in a low signal-to-noise ratio (S/N) for the obtained spectra. We also added RV data from \cite{NELSON2022101770} to our analysis to have better results (10 measurements).

We used a full Keplerian model to analyze the RV curve \citep{irwin-rv}, where the radial velocities of the components are given as
\begin{equation}
\begin{aligned}
    V_1 = V_\gamma + K_1\left( \cos(\nu+\omega)+e\cos\omega \right) \\
    V_2 = V_\gamma - K_2\left( \cos(\nu+\omega)+e\cos\omega \right) 
\end{aligned}
\end{equation}
where $V_\gamma$ is the velocity of the barycenter of the system, $K_{1,2}$ the amplitudes of the RV curves for the individual component, $e$ the orbital eccentricity, $\nu$ the true anomaly, and $\omega$ the argument of the pericenter. An additional input was the linear ephemeris (period $P$ and reference time $T_0$) determined from the TESS data. We used the differential evolution algorithm of the \textsc{scipy} package \citep{diffevol,scipy} to obtain initial parameter values. The final values of the parameters with the uncertainties estimations are a result of Monte Carlo sampling using the \textsc{emcee} package \citep{emcee}. As a next step, we derive additional parameters of the system: the total projected semi-major axis $a\sin i$ and the mass ratio $q \equiv M_2/M_1$.

We fixed the orbital eccentricity to 0. In such a case, the parameter $\omega$ has only the meaning of a phase shift between the LC and RV data and cannot be interpreted as an argument of the pericenter which is undefined for circular orbit. Testing runs with eccentricity as a fitted parameter yielded a value of nearly 0, but slightly increased the errors of the other fitted parameters. All RV measurements with the theoretical model are shown in Fig.~\ref{fig:rv}. The results of the RV analysis are listed in the upper part of Tab.~\ref{tab:photometric_parameters}.

\section{LC Analysis}

For the analysis of the light curves, we used the ELISa code \citep{2021A&A...652A.156C}, a Python package to model eclipsing binaries that includes surface features such as stellar spots.

For initial solutions, we used the Levenberg-Marquardt least-square algorithm, while the final parameters and their uncertainties were determined via Monte Carlo Markov chain (MCMC) sampling. As input data set, we initially used TESS data from Sector 57. Since the presence of a spot had not yet been considered, we selected the first of the two phase curves in \cref{FIG:phasecurves}. At this stage, our primary goal was to determine the system's fundamental parameters. The model contains four free parameters: orbital inclination \textit{i}, surface potentials of both components $\Omega_1$ and $\Omega_2$, and the effective temperature of the secondary component $T_2$. The mass ratio $q$ was fixed based on the RV analysis.

We estimated the effective temperature of the primary component $T_1$ using its mass determined from RV fitting and assuming empirical relations for the main-sequence stars \citep{Eker2018}. Its value (6750~K) is consistent with the value used by \cite{Li_2014} and is in good agreement with the expected spectral type and the absolute parameters determined in the following analysis.
We also attempted to estimate the effective temperature using our multi-color photometry and color indices \citep{Sekiguchi2000}. The resulting temperature was significantly lower than expected from the spectroscopic and photometric solution. One reason is that the observed color indices represent the combined light of both binary components. 
In addition, the system is affected by moderate interstellar reddening. Using the interstellar dust map \citep{Green2019}, we obtained reddening of $E(B-V)=0.080$ and $A_V=0.264$ mag. The primary temperature estimated from the $B-V = 0.56$ color index is about 6000~K (taking into account interstellar reddening).

\begin{figure*}
\centering
\includegraphics[width=0.49\textwidth]{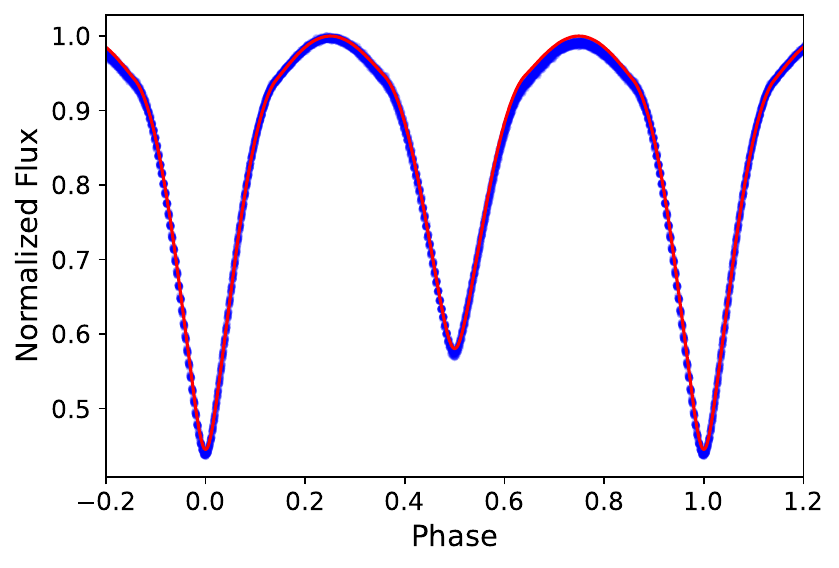}
\includegraphics[width=0.49\textwidth]{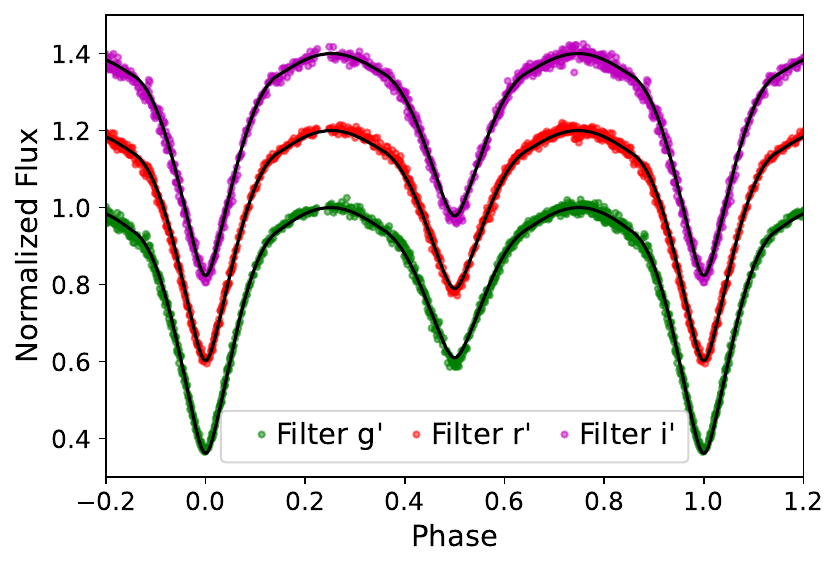}
\caption{Fits of the phase curves of the IR Cas system from the TESS satellite (\textit{left}) and from ground-based observations (\textit{right}).}
\label{fig:lc-fit}
\end{figure*}

The system parameters derived from the spectroscopic measurements, as well as the absolute parameters obtained from the analysis of both TESS and ground-based observation data, as well as the data collected from \citet{Li_2014} and \citet{NELSON2022101770} are listed in \cref{tab:photometric_parameters}. Phase curves and the corresponding models from the TESS satellite and ground-based observations are shown in \cref{fig:lc-fit}.

\begin{table*}
\centering
\caption{Parameters of the system obtained by our analysis and collected from \citet{Li_2014} and \citet{NELSON2022101770}. The upper panel lists parameters derived from spectroscopic measurements. The middle panel contains parameters obtained from the photometric solution of the light curves. The 3$^{\rm rd}$ panel presents the absolute parameters of the system. Parameters of the spot on the secondary component are listed in the last panel.}
\label{tab:photometric_parameters}
\begin{tabular}{lcccc}
\hline
Parameter & TESS & $g' + r' + i'$ & \citet{Li_2014} & \citet{NELSON2022101770} \\ \hline

$V_\gamma$ (km/s) & \multicolumn{2}{c}{$-41.85 \pm 0.68$} & --- & $-47.6 \pm 1.3$ \\
$K_1$ (km/s)      & \multicolumn{2}{c}{$142.987 \pm 1.002$} & --- & $151.9 \pm 1.7$ \\
$K_2$ (km/s)      & \multicolumn{2}{c}{$178.795 \pm 1.191$} & --- & $177.6 \pm 2.3$ \\
$e$               & \multicolumn{2}{c}{$0^a$} & --- & --- \\
$\omega$ (deg)         & \multicolumn{2}{c}{$91.218 \pm 0.373$} & --- & --- \\
$q$               & \multicolumn{2}{c}{$0.7998 \pm 0.0079$} & $0.851 \pm 0.005$ & $0.854 \pm 0.031$ \\
$a \sin i$ ($R_{\odot}$) & \multicolumn{2}{c}{$4.329 \pm 0.020$} & --- & --- \\ \hline

$\Omega_1$ & $3.468^{+0.0469}_{-0.0072}$ & $3.528^{+0.035}_{-0.020}$ & $3.504^{a}$ & $3.524 \pm 0.021$ \\
$\Omega_2$ & $3.4788^{+0.0426}_{-0.0097}$ & $3.435^{+0.1125}_{-0.0098}$ & $3.760 \pm 0.0145$ & $3.528 \pm 0.015$\\
$\Omega_{\mathrm{C}}$ & $3.40392^{+0.01347}_{-0.00082}$ & $3.4212^{+0.0065}_{-0.0061}$ & --- & --- \\
$T_1$ (K) & \multicolumn{2}{c}{$6750^{a}$} & $6750^{a}$ & ${6640^{a}}$ \\
$T_2$ (K) & $6050.2^{+1.6}_{-14.7}$ & $5942.1^{+7.4}_{-25.9}$ & $5992 \pm 4$ & $5798 \pm 6$ \\
$i$ (deg) & $85.14^{+0.16}_{-0.23}$ & $85.57^{+0.29}_{-0.20}$ & $86.8 \pm 0.2$ & $85.24 \pm 0.07$ \\ \hline

$a$ ($R_{\odot}$) & $4.345^{+0.028}_{-0.031}$ & $4.342^{+0.031}_{-0.028}$ & $4.51$ & $4.47 \pm 0.08 $ \\ 
$R_{1}^{\mathrm{eq}}$ ($R_{\odot}$) & $1.6906^{+0.0056}_{-0.0282}$ & $1.6553^{+0.0113}_{-0.0269}$ & $1.77$ & $1.75 \pm 0.02 $\\
$R_{2}^{\mathrm{eq}}$ ($R_{\odot}$) & $1.5042^{+0.0099}_{-0.0170}$ & $1.5576^{+0.0048}_{-0.0677}$ & $1.51$ & $1.62 \pm 0.02 $\\
$M_1$ ($M_{\odot}$) & $1.3253^{+0.0011}_{-0.0053}$ & $1.3147^{+0.0054}_{-0.0023}$ & $1.43$ & $1.40 \pm 0.04$\\
$M_2$ ($M_{\odot}$) & $1.0504^{+0.00641}_{-0.00078}$ & $1.0565^{+0.0029}_{-0.0045}$ & $1.22$ & $1.20 \pm 0.03 $\\
$L_1$ ($L_{\odot}$) & $5.347^{+0.034}_{-0.136}$ & $5.125^{+0.075}_{-0.095}$ & $5.85$ & $5.11 \pm 0.09 $\\
$L_2$ ($L_{\odot}$) & $2.73^{+0.036}_{-0.105}$ & $2.725^{+0.023}_{-0.434}$ & $2.63$ & $2.56 \pm 0.16$\\ \hline

Temperature factor & 0.76$^a$ & --- & --- & $0.979 \pm 0.001$ \\
Angular radius (deg) & $19.63^{+0.11}_{-0.11}$ & --- & --- & $22 \pm 1$\\
Longitude (deg) & $311.57^{+2.55}_{-2.55}$ & --- & --- & $95 \pm 5$ \\
Co-latitude (deg) & $104.876^{+1.628}_{-1.628}$ & --- & --- & $115 \pm 5$ \\
\hline
 $^a$ Fixed value.
\end{tabular}
\end{table*}


\section{Stellar spot}

As discussed in Chapter 2, the TESS data exhibit variations in the light-curve maxima, suggesting the presence of a spot. In principle, the light curve alone does not allow for a definitive determination of whether the spot is hot or cold, nor does it constrain its exact location within the system. However, cool spots, which are associated with magnetic activity in stars that have convective envelopes, are more likely to occur on the cooler secondary component (spectral type F9–G0) than on the hotter primary (F3–F5), where such activity is expected to be significantly weaker \citep{2005LRSP....2....8B}. In addition, cool spots are generally more common than hot ones. Therefore, we assumed that there was a cool spot located on the secondary component.

\begin{figure}
	\centering
	\includegraphics[width=.9\columnwidth]{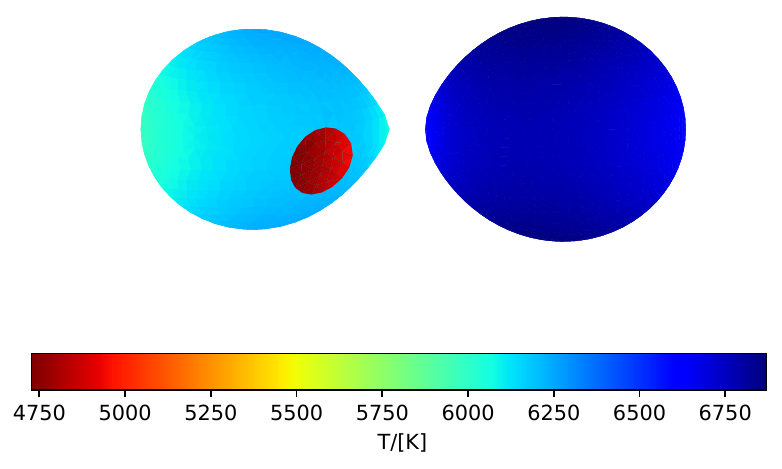}
    \includegraphics[width=1\columnwidth]{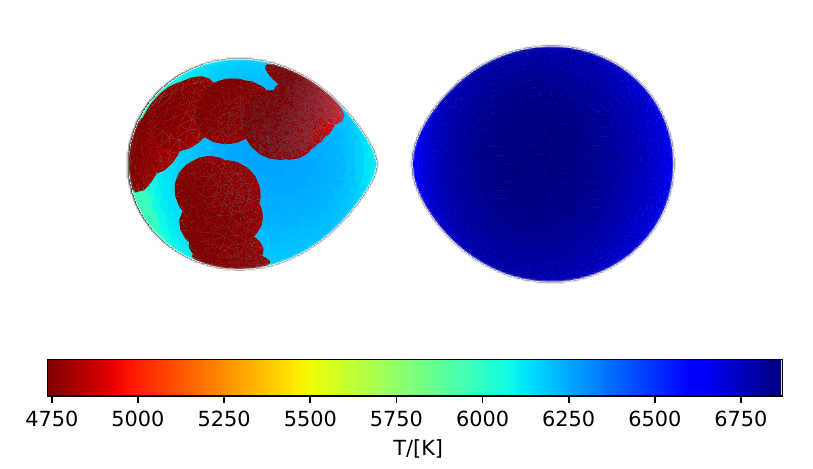}
	\caption{3D model with spot on secondary component constructed based on data from Sector 57 (\textit{top}). Spot motion across the stellar surface in polar view - all sectors combined (\textit{bottom}).}
	\label{FIG:spotmodel}
\end{figure}

To derive the fundamental parameters of the spot, we used the final segment of the light curve from Sector 57 in \cref{FIG:phasecurves}. All parameters related to the geometry of the system were fixed, and only the spot parameters (longitude, co-latitude, and angular radius) were allowed to vary. The co-latitude of the spot is equivalent to the distance from the pole, i.e. the value 90$^\circ$ means the position on the stellar equator. The spot temperature factor was 0.76, corresponding to an estimated spot temperature of approximately 4585 K \citep{2021ApJ...907...89H}.
From the initial fitting, we obtained the spot parameters (the last panel of \cref{tab:photometric_parameters}) and constructed its model (\cref{FIG:spotmodel}).


Despite having data available from only three sectors, we attempted to construct a model of the spot’s behavior on the stellar surface. The available light curves were divided into short segments, yielding a total of 64 phase curves. These were fitted individually, with the system geometry and spot size fixed, to better reveal variations in its motion.

In the initial approach, both longitude and co-latitude were allowed to vary freely, leading to abrupt, discontinuous changes in the derived spot positions between fits. After examining trends and applying appropriate limits, we removed outliers, resulting in improved consistency in the results.

The spot evolution derived from the phase-segment fitting exhibits coherent and physically acceptable behavior. 
No abrupt jumps or switches between alternative solutions are present, suggesting that the spot parameters are well constrained by the data.
The spot is located at high latitude, close to the stellar pole, which naturally results in a motion that appears as a rotation around the polar region rather than a simple longitudinal shift across the stellar surface. In this configuration, small changes in longitude result in noticeable positional shifts in the projected view, while the overall geometry remains stable.

\begin{figure}
    \centering
    \includegraphics[width=.9\columnwidth]{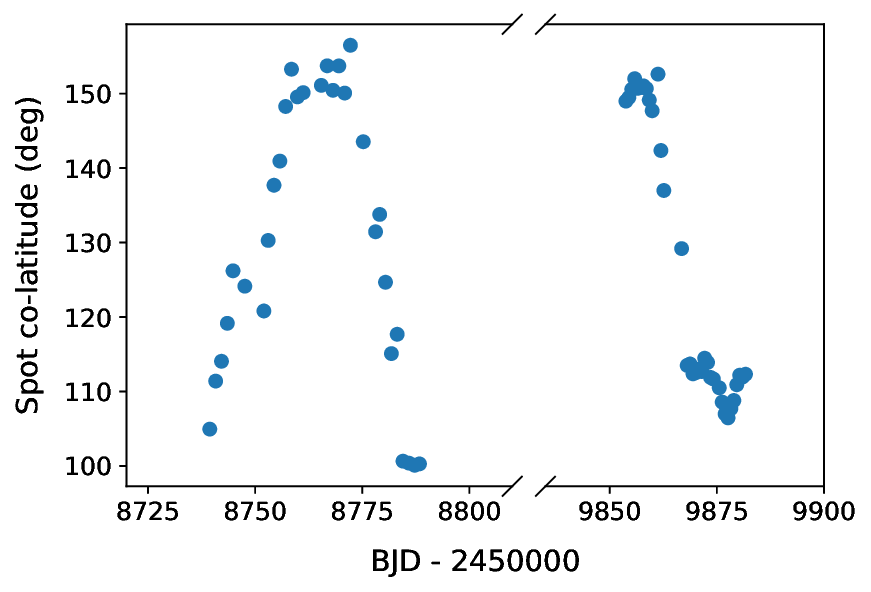}
    \caption{Time evolution of spot co-latitude.}
    \label{fig:spot-lat}
\end{figure}

The combined figure includes data from three temporally separated sectors (Fig.~\ref{FIG:spotmodel}). Consequently, discontinuities in the spot position between sectors are expected and reflect real temporal gaps rather than modeling inconsistencies. However, within each sector, the spot exhibits a consistent direction of motion (Fig.~\ref{fig:spot-lat}), supporting the interpretation of a long-lived active region.

Overall, the results indicate that spot evolution is not dominated by fitting degeneracies, but instead reflects a stable, coherent surface structure whose apparent motion is determined by stellar rotation and projection effects. Therefore, the observed morphological changes in the LC can be explained by the variation in the position of the cool starspot presented on the secondary component of this EB.

\section{Results and Discussion}

In our study, we present an analysis of a detached eclipsing binary IR Cas. We used data from \textit{TESS} space missions and  in addition, we performed ground-based photometry using Sloan filters $g'$, $r'$ and $i'$, as well as radial velocity measurements.

The O$-$C analysis based on both TESS data and historical times of minima revealed the presence of long-term quasi-periodic variations. The observed behavior can be reasonably described by the light-time effect caused by a possible third body orbiting the eclipsing pair with a period of approximately 38 years. The resulting parameters suggest a low-mass companion with a minimum mass close to $0.25\,M_{\odot}$, corresponding to a probable late-type main-sequence M-type star.

Spectroscopic analysis provided new radial velocity curves for both components and enabled an improved determination of the mass ratio and orbital parameters. The light-curve modeling successfully reproduced both the TESS and ground-based photometric observations. The system has the nearly edge-on geometry (orbital inclination about $85^\circ$) expected for a deeply eclipsing detached binary. The derived masses and radii indicate that the primary component is slightly more massive and larger than the secondary one, corresponding to late F- and early G-type main-sequence objects.

Since the TESS data exhibited asymmetries in the maxima, a cool spot located on the secondary component was introduced into the model. The obtained solution suggests the presence of a relatively stable active region whose longitudinal position changes slowly with time. Although the exact spot configuration cannot be uniquely determined, the proposed model provides a physically plausible explanation of the observed light-curve morphology.

The absolute parameters derived in this work (\cref{tab:photometric_parameters}) are generally consistent with previous studies by \cite{Li_2014} and \cite{NELSON2022101770}, although there are small differences. In particular, our solution yields slightly lower stellar masses and somewhat larger radii compared to earlier determinations. Nevertheless, all obtained parameters remain within the expected range for detached main-sequence binaries of similar spectral type.

\begin{figure}
\centering
\includegraphics[width=\columnwidth]{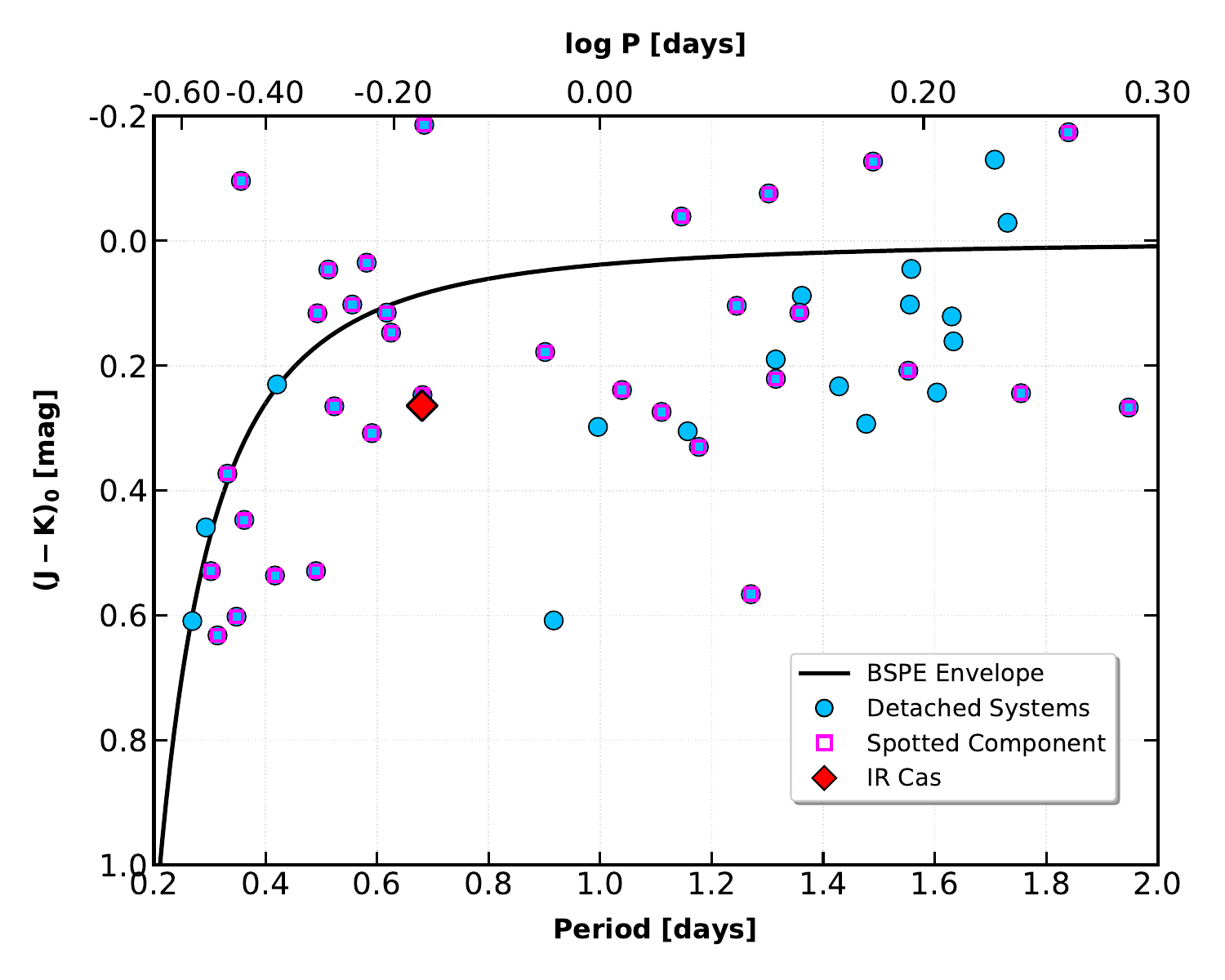}
\caption{Period--color distribution for the sample of 102 binary
systems \citep{Vanko} together with the position of the IR Cas components. The solid black line defines the BSPE for contact systems.}
\label{fig:period_color_jk}
\end{figure}

\begin{figure}
\centering
\includegraphics[width=\columnwidth]{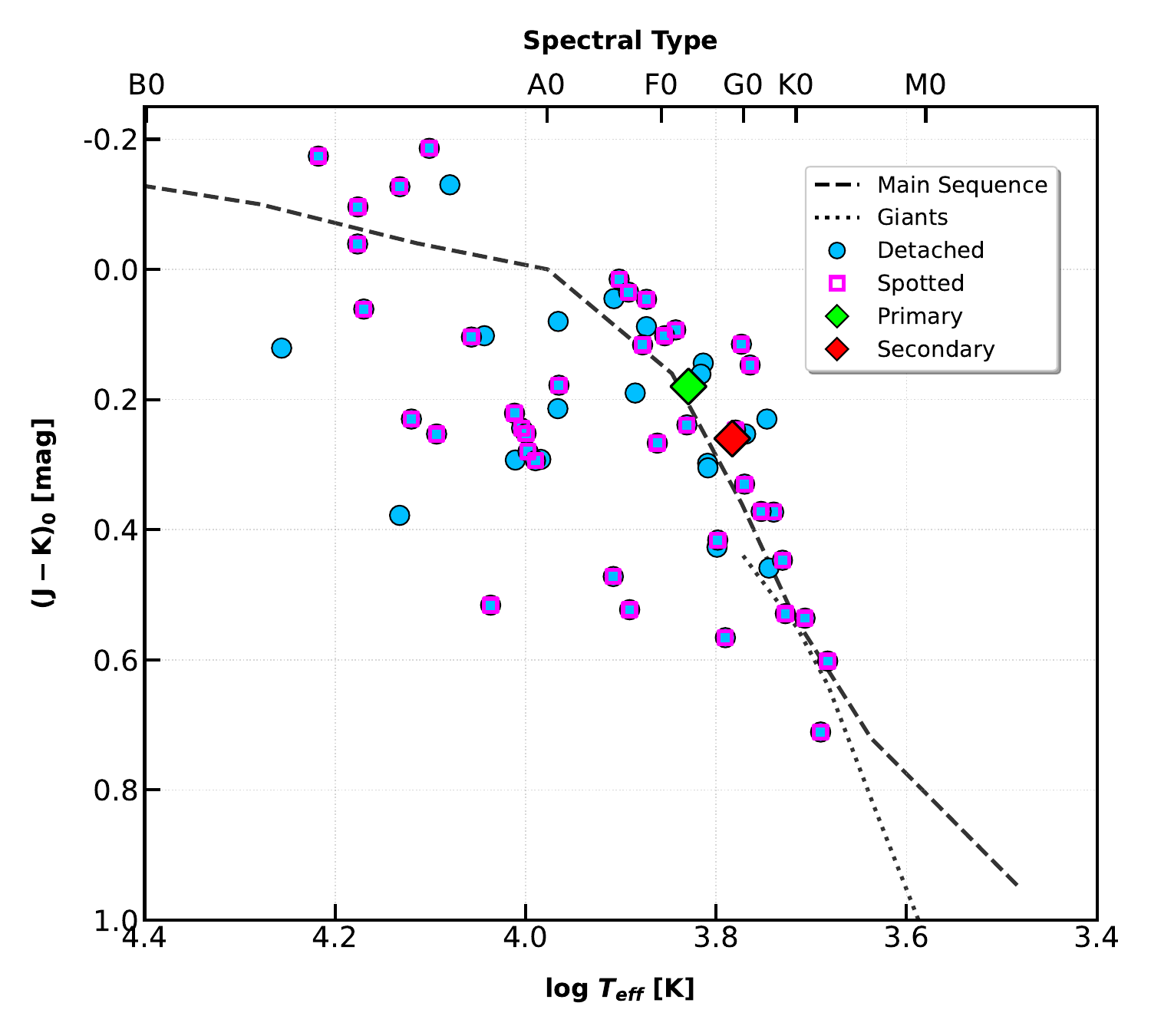}
\caption{
Relationship between effective temperature ($\log T_{\rm
eff}$) and intrinsic infrared color $(J - K)_0$ for the binary star
sample of \citet{Vanko}. The primary and secondary components of IR Cas are denoted by green and
red diamond symbols, respectively. The dashed line represents the theoretical main-sequence (Luminosity Class V) track, while the dotted line indicates the giant (Luminosity Class III) track.}
\label{RM_IR_Cas_J-K_T}
\end{figure}

\begin{figure}
\centering
\includegraphics[width=1.0\linewidth]{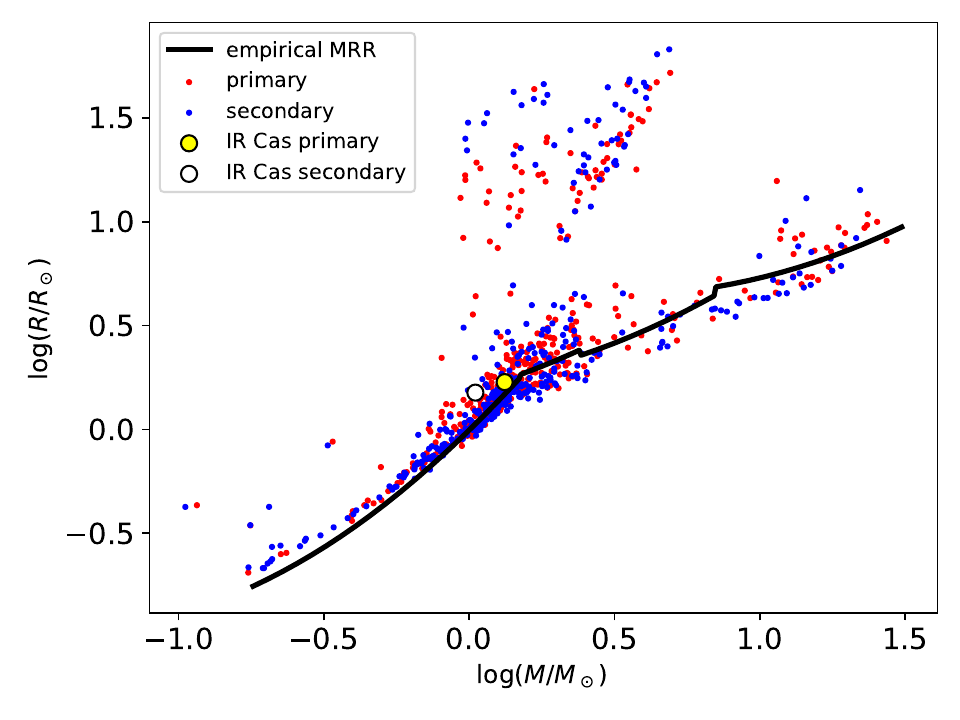}
\caption{The mass--radius 
diagram for primary (yellow circle) and secondary component (white circle) of IR Cas. Red and blue dots show similar EBs from DEBCat catalogue. The solid line represents the empirical relation for main sequence stars.}
\label{fig:TL}
\end{figure}

The position of IR\,Cas in the period--color diagram (Fig. \ref{fig:period_color_jk}) provides a concise diagnostic of its evolutionary and physical state in comparison with the sample of similar binary systems from the paper of \citet{Vanko}. Relative to the Blue Short-Period Envelope (BSPE), defined as $(J-K)_0 = 0.038P^{-2.1}$, its location indicates whether it belongs to the hot boundary of contact systems or is shifted toward cooler temperatures. This relation is adapted from the optical envelope $(V - I)$—$(V - I)_{BSP} = 0.053P^{-2.1}$—through standard color-color transformations. The $\log P$ scale  of this figure is consistent with the functional form of absolute-magnitude calibrations by \cite{Rucinski1997}. A placement near the BSPE would suggest a relatively unevolved, thermally efficient contact configuration, whereas a redward offset implies cooler components, possibly due to evolutionary effects or magnetic activity. In comparison with the overall sample, the alignment of IR Cas with either detached or contact systems further constrains its classification. Any displacement toward higher $(J-K)_0$ may also reflect the influence of stellar spots. All color indices have been corrected for interstellar reddening ($E_{J-K}$) using the neutral hydrogen column density ($N_{HI}$) method to ensure the representation of intrinsic stellar temperatures.

The location of IR\,Cas in the temperature--color diagram (Fig. \ref{RM_IR_Cas_J-K_T}) offers a direct probe of the thermal properties of its components. The primary and secondary components are plotted relative to the theoretical main-sequence and giant loci \citep{Tokunaga2000}, allowing an immediate assessment of their evolutionary status. Both components are expected to lie close to the main-sequence track, indicating that their temperatures and intrinsic colors are broadly consistent with dwarf stars. The primary, positioned toward higher $\log T_{\rm eff}$, appears hotter and bluer, while the secondary is shifted to lower temperatures and redder $(J-K)$, reflecting the typical temperature gradient within the eclipsing binaries. Any offset of either component above the main-sequence relation (toward redder colors at a given temperature) may signal magnetic activity, such as cool starspots, consistent with the behavior observed in other systems marked by spot indicators. The separation between the two components further supports a configuration with non-identical surface temperatures, as expected for many close binaries. Overall, the placement of IR Cas confirms that both stars retain main-sequence-like characteristics, with modest deviations likely attributable to activity or binary interaction effects rather than advanced evolutionary status.

We constructed the mass--radius diagram (Fig.\ref{fig:TL}) to compare IR Cas to similar detached binaries from the DEBCat sample \citep{southworth2014debcatdetachedeclipsingbinary}. Moreover, we calculated the empirical relations adopted from \citet{Eker2018} that represent the typical behaviour of main-sequence stars. Such a diagram provides a useful comparison between the physical properties of the components of IR Cas and a population of well-studied detached eclipsing binaries.

Both components of IR Cas are located very close to the empirical main sequence and fit well within the main distribution of detached systems from the DEBCat catalog. Their positions indicate that the parameters of both stars are consistent with those expected for standard main-sequence stars of comparable masses. The absence of significant deviations from the empirical relation suggests that the system has not undergone substantial mass transfer or strong mutual interaction. Furthermore, no noticeable radius inflation of the secondary component is observed, as is often seen in some interacting or magnetically active binaries. The location of both components of the IR Cas in this diagram therefore supports the interpretation that the system represents an evolutionarily representative detached binary whose components preserve the physical properties characteristic of main-sequence stars.

\begin{acknowledgements}
This article is based on the data collected with the Perek's 2-m telescope at the Astronomical Institute of the Czech Academy of Sciences in Ond\v{r}ejov.
The Slovak Research and Development Agency supported this work under contract no APVV-24-0160. This work has also been supported by the VEGA grant of the Slovak Academy of Sciences No. 2/0033/26. The research of P.G. was supported by the internal grant No. VVGS-2023-2784 of the P. J. {\v S}af{\'a}rik University in Ko{\v s}ice and funded by the EU NextGenerationEU through the Recovery and Resilience Plan for Slovakia under project No. 09I03-03-V05-00008.
\end{acknowledgements}

\facilities{OO:2, TESS}

\software{Muniwin \citep{2014ascl.soft02006H}, ELISa \citep{2021A&A...652A.156C}, iSpec \citep{ispec2014,ispec2019}, lightkurve \citep{2018ascl.soft12013L}, OCFit \citep{ocfit}}

\bibliographystyle{aasjournal}
\bibliography{refs}

@ARTICLE{1943AN....274...36H,
       author = {{Hoffmeister}, Cuno},
        title = "{213 neue Ver{\"a}nderliche}",
      journal = {Astronomische Nachrichten},
         year = 1943,
        month = jul,
       volume = {274},
        pages = {36},
          doi = {10.1002/asna.19432740109},
       adsurl = {https://ui.adsabs.harvard.edu/abs/1943AN....274...36H}
}

@ARTICLE{2006SASS...25...47W,
       author = {{Watson}, C.~L. and {Henden}, A.~A. and {Price}, A.},
        title = "{The International Variable Star Index (VSX)}",
      journal = {Society for Astronomical Sciences Annual Symposium},
         year = 2006,
        month = may,
       volume = {25},
        pages = {47},
       adsurl = {https://ui.adsabs.harvard.edu/abs/2006SASS...25...47W}
}

@article{Kirk_2016,
doi = {10.3847/0004-6256/151/3/68},
url = {https://doi.org/10.3847/0004-6256/151/3/68},
year = {2016},
month = {feb},
publisher = {The American Astronomical Society},
volume = {151},
number = {3},
pages = {68},
author = {Kirk, Brian and Conroy, Kyle and Prša, Andrej and Abdul-Masih, Michael and Kochoska, Angela and Matijevič, Gal and Hambleton, Kelly and Barclay, Thomas and Bloemen, Steven and Boyajian, Tabetha and Doyle, Laurance R. and Fulton, B. J. and Hoekstra, Abe Johannes and Jek, Kian and Kane, Stephen R. and Kostov, Veselin and Latham, David and Mazeh, Tsevi and Orosz, Jerome A. and Pepper, Joshua and Quarles, Billy and Ragozzine, Darin and Shporer, Avi and Southworth, John and Stassun, Keivan and Thompson, Susan E. and Welsh, William F. and Agol, Eric and Derekas, Aliz and Devor, Jonathan and Fischer, Debra and Green, Gregory and Gropp, Jeff and Jacobs, Tom and Johnston, Cole and LaCourse, Daryll Matthew and Saetre, Kristian and Schwengeler, Hans and Toczyski, Jacek and Werner, Griffin and Garrett, Matthew and Gore, Joanna and Martinez, Arturo O. and Spitzer, Isaac and Stevick, Justin and Thomadis, Pantelis C. and Vrijmoet, Eliot Halley and Yenawine, Mitchell and Batalha, Natalie and Borucki, William},
title = {KEPLER ECLIPSING BINARY STARS. VII. THE CATALOG OF ECLIPSING BINARIES FOUND IN THE ENTIRE KEPLER DATA SET},
journal = {The Astronomical Journal}
}

@article{Kostov_2025,
doi = {10.3847/1538-4365/ade2d8},
url = {https://doi.org/10.3847/1538-4365/ade2d8},
year = {2025},
month = {aug},
publisher = {The American Astronomical Society},
volume = {279},
number = {2},
pages = {50},
author = {Kostov, Veselin B. and Powell, Brian P. and Fornear, Aline U. and Di Fraia, Marco Z. and Gagliano, Robert and Jacobs, Thomas L. and de Lambilly, Julien S. and Durantini Luca, Hugo A. and Majewski, Steven R. and Omohundro, Mark and Orosz, Jerome and Rappaport, Saul A. and Salik, Ryan and Short, Donald and Welsh, William and Alexandrov, Svetoslav and da Silva, Cledison Marcos and Dunning, Erika and Gühne, Gerd and Huten, Marc and Hyogo, Michiharu and Iannone, Davide and Lee, Sam and Magliano, Christian and Sharma, Manya and Tarr, Allan and Yablonsky, John and Acharya, Sovan and Adams, Fred and Barclay, Thomas and Montet, Benjamin T. and Mullally, Susan and Olmschenk, Greg and Prša, Andrej and Quintana, Elisa and Wilson, Robert and Balcioglu, Hasret and Kruse, Ethan and The Eclipsing Binary Patrol Collaboration},
title = {The TESS Ten Thousand Catalog: 10,001 Uniformly Vetted and Validated Eclipsing Binary Stars Detected in Full-frame Image Data by Machine Learning and Analyzed by Citizen Scientists},
journal = {The Astrophysical Journal Supplement Series},
}

@article{Mowlavi_2023,
   title={GaiaData Release 3: The first Gaia catalogue of eclipsing-binary candidates},
   volume={674},
   ISSN={1432-0746},
   url={http://dx.doi.org/10.1051/0004-6361/202245330},
   DOI={10.1051/0004-6361/202245330},
   journal={\aap},
   publisher={EDP Sciences},
   author={Mowlavi, N. and Holl, B. and Lecoeur-Taïbi, I. and Barblan, F. and Kochoska, A. and Prša, A. and Mazeh, T. and Rimoldini, L. and Gavras, P. and Audard, M. and Jevardat de Fombelle, G. and Nienartowicz, K. and García-Lario, P. and Eyer, L.},
   year={2023},
   month=jun, pages={A16} }

@software{2018ascl.soft12013L,
       author = {{Lightkurve Collaboration}},
        title = "{Lightkurve: Kepler and TESS time series analysis in Python}",
 howpublished = {Astrophysics Source Code Library, record ascl:1812.013},
         year = 2018,
        month = dec,
          eid = {ascl:1812.013},
archivePrefix = {ascl},
       eprint = {1812.013},
       adsurl = {https://ui.adsabs.harvard.edu/abs/2018ascl.soft12013L}
}

@ARTICLE{1978IBVS.1446....1K,
       author = {{Kreiner}, J.~M. and {Tremko}, J.},
        title = "{Peculiarities of Some beta Lyrae-Type Stars and the Need of Their Further Investigation}",
      journal = {Information Bulletin on Variable Stars},
         year = 1978,
        month = jul,
       volume = {1446},
        pages = {1},
       adsurl = {https://ui.adsabs.harvard.edu/abs/1978IBVS.1446....1K}
}

@article{Li_2014,
   title={THE FIRST PHOTOMETRIC ANALYSIS OF THE NEAR CONTACT BINARY IR Cas},
   volume={148},
   ISSN={1538-3881},
   url={http://dx.doi.org/10.1088/0004-6256/148/5/96},
   DOI={10.1088/0004-6256/148/5/96},
   number={5},
   journal={The Astronomical Journal},
   publisher={American Astronomical Society},
   author={Li, Kai and Hu, S.-M. and Guo, D.-F. and Jiang, Y.-G. and Gao, D.-Y. and Chen, X.},
   year={2014},
   month=oct, pages={96} }

@article{NELSON2022101770,
title = {The detached binary IR cassiopeiae},
journal = {New Astronomy},
volume = {93},
pages = {101770},
year = {2022},
issn = {1384-1076},
doi = {https://doi.org/10.1016/j.newast.2022.101770},
url = {https://www.sciencedirect.com/science/article/pii/S1384107622000045},
author = {Robert H. Nelson},
}

@ARTICLE{2021A&A...652A.156C,
       author = {{\v{C}okina}, Michal and {Fedurco}, Miroslav and {Parimucha}, {\v{S}}tefan},
        title = "{ELISa: A new tool for fast modelling of eclipsing binaries}",
      journal = {\aap},
         year = 2021,
        month = aug,
       volume = {652},
          eid = {A156},
        pages = {A156},
          doi = {10.1051/0004-6361/202039171},
archivePrefix = {arXiv},
       eprint = {2106.10116},
 primaryClass = {astro-ph.IM},
       adsurl = {https://ui.adsabs.harvard.edu/abs/2021A&A...652A.156C}
}

@article{oes1,
author = {Koubsky, P. and Mayer, P. and Čáp, Jiří and Zdárský, F. and Zeman, J. and Pina, Ladislav and Melich, Z.},
year = {2004},
month = {01},
pages = {},
title = {Ondrejov Echelle Spectrograph - OES},
journal = {Publications of the Astronomical Institute of the Czechoslovak Academy of Sciences}
}

@ARTICLE{oes2,
       author = {{Kab{\'a}th}, P. and {Skarka}, M. and {Sabotta}, S. and {Guenther}, E. and {Jones}, D. and {Klocov{\'a}}, T. and {{\v{S}}ubjak}, J. and {{\v{Z}}{\'a}k}, Ji{\v{r}}{\'\i} and {{\v{S}}pokov{\'a}}, M. and {Bla{\v{z}}ek}, M. and {Dvo{\v{r}}{\'a}kov{\'a}}, J. and {Dupkala}, D. and {Fuchs}, J. and {Hatzes}, A. and {Kortusov{\'a}}, E. and {Novotn{\'y}}, R. and {Pl{\'a}valov{\'a}}, E. and {{\v{R}}ezba}, L. and {Sloup}, J. and {{\v{S}}koda}, P. and {{\v{S}}lechta}, M.},
        title = "{Ond{\v{r}}ejov Echelle Spectrograph, Ground Based Support Facility for Exoplanet Missions}",
      journal = {\pasp},
         year = 2020,
        month = mar,
       volume = {132},
       number = {1009},
          eid = {035002},
        pages = {035002},
          doi = {10.1088/1538-3873/ab6752},
archivePrefix = {arXiv},
       eprint = {2001.01001},
 primaryClass = {astro-ph.IM},
       adsurl = {https://ui.adsabs.harvard.edu/abs/2020PASP..132c5002K}
}

@ARTICLE{musikos,
       author = {{Pribulla}, T. and {Va{\v{n}}ko}, M. and {Kom{\v{z}}{\'\i}k}, R. and {Sivani{\v{c}}}, P.},
        title = "{High-resolution {\'e}chelle spectrograph at Skalnat{\'e} Pleso Observatory}",
      journal = {Contributions of the Astronomical Observatory Skalnate Pleso},
         year = 2024,
        month = feb,
       volume = {54},
       number = {2},
        pages = {43-46},
          doi = {10.31577/caosp.2024.54.2.43},
       adsurl = {https://ui.adsabs.harvard.edu/abs/2024CoSka..54b..43P}
}

@ARTICLE{Griffin1967,
       author = {{Griffin}, R.~F.},
        title = "{A Photoelectric Radial-Velocity Spectrometer}",
      journal = {\apj},
         year = 1967,
        month = may,
       volume = {148},
        pages = {465},
          doi = {10.1086/149168},
       adsurl = {https://ui.adsabs.harvard.edu/abs/1967ApJ...148..465G}
}

@ARTICLE{Simkin1974,
       author = {{Simkin}, S.~M.},
        title = "{Measurements of Velocity Dispersions and Doppler Shifts from Digitized Optical Spectra}",
      journal = {\aap},
         year = 1974,
        month = mar,
       volume = {31},
        pages = {129},
       adsurl = {https://ui.adsabs.harvard.edu/abs/1974A&A....31..129S}
}

@ARTICLE{TonryDavis1979,
       author = {{Tonry}, J. and {Davis}, M.},
        title = "{A survey of galaxy redshifts. I. Data reduction techniques.}",
      journal = {\aj},
         year = 1979,
        month = oct,
       volume = {84},
        pages = {1511-1525},
          doi = {10.1086/112569},
       adsurl = {https://ui.adsabs.harvard.edu/abs/1979AJ.....84.1511T}
}

@ARTICLE{Zverko2007,
       author = {{Zverko}, J. and {{\v{Z}}i{\v{z}}{\v{n}}ovsk{\'y}}, J. and {Mikul{\'a}{\v{s}}ek}, Z. and {Iliev}, I. Kh.},
        title = "{Radial velocity determination by CCF using a synthetic spectrum as the template and detecting component spectra in SB1 binaries}",
      journal = {Contributions of the Astronomical Observatory Skalnate Pleso},
         year = 2007,
        month = feb,
       volume = {37},
       number = {1},
        pages = {49-62},
       adsurl = {https://ui.adsabs.harvard.edu/abs/2007CoSka..37...49Z}
}

@ARTICLE{Pribulla2015,
       author = {{Pribulla}, T. and {Garai}, Z. and {Hamb{\'a}lek}, L. and {Koll{\'a}r}, V. and {Kom{\v{z}}{\'\i}k}, R. and {Kundra}, E. and {Nedoro{\v{s}}{\v{c}}{\'\i}k}, J. and {Seker{\'a}{\v{s}}}, M. and {Va{\v{n}}ko}, M},
        title = "{Affordable {\'e}chelle spectroscopy with a 60 cm telescope}",
      journal = {Astronomische Nachrichten},
         year = 2015,
        month = sep,
       volume = {336},
       number = {7},
        pages = {682},
          doi = {10.1002/asna.201512202},
       adsurl = {https://ui.adsabs.harvard.edu/abs/2015AN....336..682P}
}

@ARTICLE{Garai2017,
       author = {{Garai}, Z. and {Pribulla}, T. and {Hamb{\'a}lek}, {\L}. and {Kundra}, E. and {Va{\v{n}}ko}, M. and {Raetz}, S. and {Seeliger}, M. and {Marka}, C. and {Gilbert}, H.},
        title = "{Affordable echelle spectroscopy of the eccentric HAT-P-2, WASP-14, and XO-3 planetary systems with a sub-meter-class telescope}",
      journal = {Astronomische Nachrichten},
         year = 2017,
        month = jan,
       volume = {338},
       number = {1},
        pages = {35-48},
          doi = {10.1002/asna.201613208},
archivePrefix = {arXiv},
       eprint = {1608.00745},
 primaryClass = {astro-ph.IM},
       adsurl = {https://ui.adsabs.harvard.edu/abs/2017AN....338...35G}
}

@ARTICLE{Pych2004,
       author = {{Pych}, Wojtek},
        title = "{A Fast Algorithm for Cosmic-Ray Removal from Single Images}",
      journal = {\pasp},
         year = 2004,
        month = feb,
       volume = {116},
       number = {816},
        pages = {148-153},
          doi = {10.1086/381786},
archivePrefix = {arXiv},
       eprint = {astro-ph/0311290},
 primaryClass = {astro-ph},
       adsurl = {https://ui.adsabs.harvard.edu/abs/2004PASP..116..148P}
}

@INPROCEEDINGS{Tody1986,
       author = {{Tody}, Doug},
        title = "{The IRAF Data Reduction and Analysis System}",
    booktitle = {Instrumentation in astronomy VI},
         year = 1986,
       editor = {{Crawford}, David L.},
       series = {Society of Photo-Optical Instrumentation Engineers (SPIE) Conference Series},
       volume = {627},
        month = jan,
        pages = {733},
          doi = {10.1117/12.968154},
       adsurl = {https://ui.adsabs.harvard.edu/abs/1986SPIE..627..733T}
}

@ARTICLE{2021ApJ...907...89H,
       author = {{Herbst}, Konstantin and {Papaioannou}, Athanasios and {Airapetian}, Vladimir S. and {Atri}, Dimitra},
        title = "{From Starspots to Stellar Coronal Mass Ejections{\textemdash}Revisiting Empirical Stellar Relations}",
      journal = {\apj},
         year = 2021,
        month = feb,
       volume = {907},
       number = {2},
          eid = {89},
        pages = {89},
          doi = {10.3847/1538-4357/abcc04},
archivePrefix = {arXiv},
       eprint = {2011.03761},
 primaryClass = {astro-ph.SR},
       adsurl = {https://ui.adsabs.harvard.edu/abs/2021ApJ...907...89H}
}

@ARTICLE{2005LRSP....2....8B,
       author = {{Berdyugina}, Svetlana V.},
        title = "{Starspots: A Key to the Stellar Dynamo}",
      journal = {Living Reviews in Solar Physics},
         year = 2005,
        month = dec,
       volume = {2},
       number = {1},
          eid = {8},
        pages = {8},
          doi = {10.12942/lrsp-2005-8},
       adsurl = {https://ui.adsabs.harvard.edu/abs/2005LRSP....2....8B}
}

@INBOOK{Tokunaga2000,
       author = {{Tokunaga}, A.~T.},
        title = "{Stellar Parameters}",
     booktitle = {Allen's Astrophysical Quantities},
         year = 2000,
       editor = {{Cox}, Arthur N.},
      edition = {4th},
    publisher = {Springer-Verlag},
      address = {New York},
        pages = {143},
       adsurl = {https://ui.adsabs.harvard.edu/abs/2000asqu.book..143T}
      
}

@ARTICLE{Rucinski1997,
       author = {{Rucinski}, Slavek M. and {Duerbeck}, Hilmar W.},
        title = "{Absolute-Magnitude Calibration for the W Ursae Majoris-Type Systems Based on HIPPARCOS Data}",
      journal = {Publications of the Astronomical Society of the Pacific},
         year = 1997,
        month = dec,
       volume = {109},
        pages = {1340-1350},
          doi = {10.1086/134014},
       adsurl = {https://ui.adsabs.harvard.edu/abs/1997PASP..109.1340R}

}

@article{Vanko,
  author  = {{Va\v{n}ko}, Martin and {Kamenec}, Mat\'u\v{s} and {Gajdo\v{s}}, Pavol and Pribulla, Theodor and Munagala, Reddy Charan Reddy and {Parimucha}, {\v{S}}tefan},
  title   = "{Evolutionary status and classification of selected neglected eclipsing binaries observed with TESS}",
  journal = {\aj},
  volume    = {submitted},
  year    = {2026}

  }

@ARTICLE{Eker2018,
       author = {{Eker}, Z. and {Bak{\i}{\textcommabelow s}}, V. and {Bilir}, S. and {Soydugan}, F. and {Steer}, I. and {Soydugan}, E. and {Bak{\i}{\textcommabelow s}}, H. and {Ali{\c{c}}avu{\textcommabelow s}}, F. and {Aslan}, G. and {Alpsoy}, M.},
        title = "{Interrelated main-sequence mass-luminosity, mass-radius, and mass-effective temperature relations}",
      journal = {\mnras},
         year = 2018,
        month = oct,
       volume = {479},
       number = {4},
        pages = {5491-5511},
          doi = {10.1093/mnras/sty1834},
archivePrefix = {arXiv},
       eprint = {1807.02568},
 primaryClass = {astro-ph.SR},
       adsurl = {https://ui.adsabs.harvard.edu/abs/2018MNRAS.479.5491E}
}

@ARTICLE{Sekiguchi2000,
       author = {{Sekiguchi}, Maki and {Fukugita}, Masataka},
        title = "{A Study of the B-V Color-Temperature Relation}",
      journal = {\aj},
         year = 2000,
        month = aug,
       volume = {120},
       number = {2},
        pages = {1072-1084},
          doi = {10.1086/301490},
archivePrefix = {arXiv},
       eprint = {astro-ph/9904299},
 primaryClass = {astro-ph},
       adsurl = {https://ui.adsabs.harvard.edu/abs/2000AJ....120.1072S}
}

@book{Hilditch_2001, 
        place={Cambridge}, 
        title={An Introduction to Close Binary Stars}, 
        publisher={Cambridge University Press}, 
        author={Hilditch, R. W.}, 
        year={2001}}

@ARTICLE{2018A&A...616A...1G,
       author = {{Gaia Collaboration}},
        title = "{Gaia Data Release 2. Summary of the contents and survey properties}",
      journal = {\aap},
         year = 2018,
        month = aug,
       volume = {616},
          eid = {A1},
        pages = {A1},
          doi = {10.1051/0004-6361/201833051},
archivePrefix = {arXiv},
       eprint = {1804.09365},
 primaryClass = {astro-ph.GA},
       adsurl = {https://ui.adsabs.harvard.edu/abs/2018A&A...616A...1G}
}

@ARTICLE{2023A&A...674A...1G,
       author = {{Gaia Collaboration}},
        title = "{Gaia Data Release 3. Summary of the content and survey properties}",
      journal = {\aap},
         year = 2023,
        month = jun,
       volume = {674},
          eid = {A1},
        pages = {A1},
          doi = {10.1051/0004-6361/202243940},
archivePrefix = {arXiv},
       eprint = {2208.00211},
 primaryClass = {astro-ph.GA},
       adsurl = {https://ui.adsabs.harvard.edu/abs/2023A&A...674A...1G}
}

@ARTICLE{ocfit,
       author = {{Gajdo{\v{s}}}, P. and {Parimucha}, {\v{S}}.},
        title = "{New tool with GUI for fitting O-C diagrams}",
      journal = {Open European Journal on Variable Stars},
         year = 2019,
        month = apr,
       volume = {197},
        pages = {71},
       adsurl = {https://ui.adsabs.harvard.edu/abs/2019OEJV..197...71G}
}

@ARTICLE{2006AJ....131.1163S,
       author = {{Skrutskie}, M.~F. and {Cutri}, R.~M. and {Stiening}, R. and {Weinberg}, M.~D. and {Schneider}, S. and {Carpenter}, J.~M. and {Beichman}, C. and {Capps}, R. and {Chester}, T. and {Elias}, J. and {Huchra}, J. and {Liebert}, J. and {Lonsdale}, C. and {Monet}, D.~G. and {Price}, S. and {Seitzer}, P. and {Jarrett}, T. and {Kirkpatrick}, J.~D. and {Gizis}, J.~E. and {Howard}, E. and {Evans}, T. and {Fowler}, J. and {Fullmer}, L. and {Hurt}, R. and {Light}, R. and {Kopan}, E.~L. and {Marsh}, K.~A. and {McCallon}, H.~L. and {Tam}, R. and {Van Dyk}, S. and {Wheelock}, S.},
        title = "{The Two Micron All Sky Survey (2MASS)}",
      journal = {\aj},
         year = 2006,
        month = feb,
       volume = {131},
       number = {2},
        pages = {1163-1183},
          doi = {10.1086/498708},
       adsurl = {https://ui.adsabs.harvard.edu/abs/2006AJ....131.1163S}
}

@ARTICLE{2010AJ....140.1868W,
       author = {{Wright}, Edward L. and {Eisenhardt}, Peter R.~M. and {Mainzer}, Amy K. and {Ressler}, Michael E. and {Cutri}, Roc M. and {Jarrett}, Thomas and {Kirkpatrick}, J. Davy and {Padgett}, Deborah and {McMillan}, Robert S. and {Skrutskie}, Michael and {Stanford}, S.~A. and {Cohen}, Martin and {Walker}, Russell G. and {Mather}, John C. and {Leisawitz}, David and {Gautier}, III, Thomas N. and {McLean}, Ian and {Benford}, Dominic and {Lonsdale}, Carol J. and {Blain}, Andrew and {Mendez}, Bryan and {Irace}, William R. and {Duval}, Valerie and {Liu}, Fengchuan and {Royer}, Don and {Heinrichsen}, Ingolf and {Howard}, Joan and {Shannon}, Mark and {Kendall}, Martha and {Walsh}, Amy L. and {Larsen}, Mark and {Cardon}, Joel G. and {Schick}, Scott and {Schwalm}, Mark and {Abid}, Mohamed and {Fabinsky}, Beth and {Naes}, Larry and {Tsai}, Chao-Wei},
        title = "{The Wide-field Infrared Survey Explorer (WISE): Mission Description and Initial On-orbit Performance}",
      journal = {\aj},
         year = 2010,
        month = dec,
       volume = {140},
       number = {6},
        pages = {1868-1881},
          doi = {10.1088/0004-6256/140/6/1868},
archivePrefix = {arXiv},
       eprint = {1008.0031},
 primaryClass = {astro-ph.IM},
       adsurl = {https://ui.adsabs.harvard.edu/abs/2010AJ....140.1868W}
}

@PROCEEDINGS{1988iras....1.....B,
        title = "{Infrared Astronomical Satellite (IRAS) Catalogs and Atlases.Volume 1: Explanatory Supplement.}",
    booktitle = {Infrared astronomical satellite (IRAS) catalogs and atlases. Volume 1: Explanatory supplement},
         year = 1988,
       editor = {{Beichman}, C.~A. and {Neugebauer}, G. and {Habing}, H.~J. and {Clegg}, P.~E. and {Chester}, Thomas J.},
       volume = {1},
        month = jan,
       adsurl = {https://ui.adsabs.harvard.edu/abs/1988iras....1.....B}
}

@article{Gajdos_2024,
doi = {10.3847/1538-3881/ad6dd3},
url = {https://doi.org/10.3847/1538-3881/ad6dd3},
year = {2024},
month = {sep},
publisher = {The American Astronomical Society},
volume = {168},
number = {4},
pages = {171},
author = {Gajdo\v{s}, Pavol and Parimucha, {\v{S}}tefan and Skarka, Marek and Kamenec, Mat\'{u}\v{s} and Lipt\'{a}k, Jozef and Karjalainen, Raine},
title = {Analysis of KIC 7023917: Spotted Low-mass Ratio Eclipsing Binary with $\delta$ Scuti Pulsations},
journal = {The Astronomical Journal},
}

@software{2014ascl.soft02006H,
       author = {{Hroch}, Filip},
        title = "{Munipack: General astronomical image processing software}",
 howpublished = {Astrophysics Source Code Library, record ascl:1402.006},
         year = 2014,
        month = feb,
          eid = {ascl:1402.006},
archivePrefix = {ascl},
       eprint = {1402.006},
       adsurl = {https://ui.adsabs.harvard.edu/abs/2014ascl.soft02006H}
}

@INPROCEEDINGS{ocgate,
       author = {{Br{\'a}t}, L. and {Zejda}, M.},
        title = "{Variable Star and Exoplanet Section of the Czech Astronomical Society}",
    booktitle = {Binaries - Key to Comprehension of the Universe},
         year = 2010,
       editor = {{Pr{\v{s}}a}, A. and {Zejda}, M.},
       series = {Astronomical Society of the Pacific Conference Series},
       volume = {435},
        month = dec,
        pages = {457},
       adsurl = {https://ui.adsabs.harvard.edu/abs/2010ASPC..435..457B}
}

@ARTICLE{irwin-lite,
       author = {{Irwin}, John B.},
        title = "{Standard light-time curves}",
      journal = {\aj},
         year = 1959,
        month = may,
       volume = {64},
        pages = {149},
          doi = {10.1086/107913},
       adsurl = {https://ui.adsabs.harvard.edu/abs/1959AJ.....64..149I}
}

@ARTICLE{irwin-rv,
       author = {{Irwin}, John B.},
        title = "{The Determination of a Spectroscopic Binary Orbit.}",
      journal = {\apj},
         year = 1952,
        month = jul,
       volume = {116},
        pages = {218},
          doi = {10.1086/145605},
       adsurl = {https://ui.adsabs.harvard.edu/abs/1952ApJ...116..218I}
}

@software{pyraf,
       author = {{Science Software Branch at STScI}},
        title = "{PyRAF: Python alternative for IRAF}",
 howpublished = {Astrophysics Source Code Library, record ascl:1207.011},
         year = 2012,
        month = jul,
          eid = {ascl:1207.011},
archivePrefix = {ascl},
       eprint = {1207.011},
       adsurl = {https://ui.adsabs.harvard.edu/abs/2012ascl.soft07011S}
}

@ARTICLE{ispec2019,
       author = {{Blanco-Cuaresma}, Sergi},
        title = "{Modern stellar spectroscopy caveats}",
      journal = {\mnras},
         year = 2019,
        month = jun,
       volume = {486},
       number = {2},
        pages = {2075-2101},
          doi = {10.1093/mnras/stz549},
archivePrefix = {arXiv},
       eprint = {1902.09558},
 primaryClass = {astro-ph.SR},
       adsurl = {https://ui.adsabs.harvard.edu/abs/2019MNRAS.486.2075B}
}

@ARTICLE{ispec2014,
       author = {{Blanco-Cuaresma}, S. and {Soubiran}, C. and {Heiter}, U. and {Jofr{\'e}}, P.},
        title = "{Determining stellar atmospheric parameters and chemical abundances of FGK stars with iSpec}",
      journal = {\aap},
         year = 2014,
        month = sep,
       volume = {569},
          eid = {A111},
        pages = {A111},
          doi = {10.1051/0004-6361/201423945},
archivePrefix = {arXiv},
       eprint = {1407.2608},
 primaryClass = {astro-ph.IM},
       adsurl = {https://ui.adsabs.harvard.edu/abs/2014A&A...569A.111B}
}

@ARTICLE{Coelho2014,
       author = {{Coelho}, P.~R.~T.},
        title = "{A new library of theoretical stellar spectra with scaled-solar and {\ensuremath{\alpha}}-enhanced mixtures}",
      journal = {\mnras},
         year = 2014,
        month = may,
       volume = {440},
       number = {2},
        pages = {1027-1043},
          doi = {10.1093/mnras/stu365},
archivePrefix = {arXiv},
       eprint = {1404.3243},
 primaryClass = {astro-ph.SR},
       adsurl = {https://ui.adsabs.harvard.edu/abs/2014MNRAS.440.1027C}
}

@ARTICLE{emcee,
       author = {{Foreman-Mackey}, Daniel and {Hogg}, David W. and {Lang}, Dustin and {Goodman}, Jonathan},
        title = "{emcee: The MCMC Hammer}",
      journal = {\pasp},
         year = 2013,
        month = mar,
       volume = {125},
       number = {925},
        pages = {306},
          doi = {10.1086/670067},
archivePrefix = {arXiv},
       eprint = {1202.3665},
 primaryClass = {astro-ph.IM},
       adsurl = {https://ui.adsabs.harvard.edu/abs/2013PASP..125..306F}
}

@ARTICLE{diffevol,
       author = {{Storn}, Rainer and {Price}, Kenneth},
        title = "{Differential Evolution - A Simple and Efficient Heuristic for global Optimization over Continuous Spaces}",
      journal = {Journal of Global Optimization},
         year = 1997,
        month = dec,
       volume = {11},
        pages = {341-359},
          doi = {10.1023/A:1008202821328},
       adsurl = {https://ui.adsabs.harvard.edu/abs/1997JGOpt..11..341S}
}

@ARTICLE{scipy,
  author  = {Virtanen, Pauli and Gommers, Ralf and Oliphant, Travis E. and
            Haberland, Matt and Reddy, Tyler and Cournapeau, David and
            Burovski, Evgeni and Peterson, Pearu and Weckesser, Warren and
            Bright, Jonathan and {van der Walt}, St{\'e}fan J. and
            Brett, Matthew and Wilson, Joshua and Millman, K. Jarrod and
            Mayorov, Nikolay and Nelson, Andrew R. J. and Jones, Eric and
            Kern, Robert and Larson, Eric and Carey, C J and
            Polat, {\.I}lhan and Feng, Yu and Moore, Eric W. and
            {VanderPlas}, Jake and Laxalde, Denis and Perktold, Josef and
            Cimrman, Robert and Henriksen, Ian and Quintero, E. A. and
            Harris, Charles R. and Archibald, Anne M. and
            Ribeiro, Ant{\^o}nio H. and Pedregosa, Fabian and
            {van Mulbregt}, Paul and {SciPy 1.0 Contributors}},
  title   = {{{SciPy} 1.0: Fundamental Algorithms for Scientific
            Computing in Python}},
  journal = {Nature Methods},
  year    = {2020},
  volume  = {17},
  pages   = {261--272},
  adsurl  = {https://rdcu.be/b08Wh},
  doi     = {10.1038/s41592-019-0686-2},
}

@ARTICLE{2000A&AS..143....9W,
       author = {{Wenger}, M. and {Ochsenbein}, F. and {Egret}, D. and {Dubois}, P. and {Bonnarel}, F. and {Borde}, S. and {Genova}, F. and {Jasniewicz}, G. and {Lalo{\"e}}, S. and {Lesteven}, S. and {Monier}, R.},
        title = "{The SIMBAD astronomical database. The CDS reference database for astronomical objects}",
      journal = {\aaps},
         year = 2000,
        month = apr,
       volume = {143},
        pages = {9-22},
          doi = {10.1051/aas:2000332},
archivePrefix = {arXiv},
       eprint = {astro-ph/0002110},
 primaryClass = {astro-ph},
       adsurl = {https://ui.adsabs.harvard.edu/abs/2000A&AS..143....9W}
}

@INPROCEEDINGS{southworth2014debcatdetachedeclipsingbinary,
       author = {{Southworth}, J.},
        title = "{DEBCat: A Catalog of Detached Eclipsing Binary Stars}",
    booktitle = {Living Together: Planets, Host Stars and Binaries},
         year = 2015,
       editor = {{Rucinski}, S.~M. and {Torres}, G. and {Zejda}, M.},
       series = {Astronomical Society of the Pacific Conference Series},
       volume = {496},
        month = jul,
        pages = {164},
          doi = {10.48550/arXiv.1411.1219},
archivePrefix = {arXiv},
       eprint = {1411.1219},
 primaryClass = {astro-ph.SR},
       adsurl = {https://ui.adsabs.harvard.edu/abs/2015ASPC..496..164S}
}

@ARTICLE{2010Sci...327..977B,
       author = {{Borucki}, William J. and {Koch}, David and {Basri}, Gibor and {Batalha}, Natalie and {Brown}, Timothy and {Caldwell}, Douglas and {Caldwell}, John and {Christensen-Dalsgaard}, J{\o}rgen and {Cochran}, William D. and {DeVore}, Edna and {Dunham}, Edward W. and {Dupree}, Andrea K. and {Gautier}, Thomas N. and {Geary}, John C. and {Gilliland}, Ronald and {Gould}, Alan and {Howell}, Steve B. and {Jenkins}, Jon M. and {Kondo}, Yoji and {Latham}, David W. and {Marcy}, Geoffrey W. and {Meibom}, S{\o}ren and {Kjeldsen}, Hans and {Lissauer}, Jack J. and {Monet}, David G. and {Morrison}, David and {Sasselov}, Dimitar and {Tarter}, Jill and {Boss}, Alan and {Brownlee}, Don and {Owen}, Toby and {Buzasi}, Derek and {Charbonneau}, David and {Doyle}, Laurance and {Fortney}, Jonathan and {Ford}, Eric B. and {Holman}, Matthew J. and {Seager}, Sara and {Steffen}, Jason H. and {Welsh}, William F. and {Rowe}, Jason and {Anderson}, Howard and {Buchhave}, Lars and {Ciardi}, David and {Walkowicz}, Lucianne and {Sherry}, William and {Horch}, Elliott and {Isaacson}, Howard and {Everett}, Mark E. and {Fischer}, Debra and {Torres}, Guillermo and {Johnson}, John Asher and {Endl}, Michael and {MacQueen}, Phillip and {Bryson}, Stephen T. and {Dotson}, Jessie and {Haas}, Michael and {Kolodziejczak}, Jeffrey and {Van Cleve}, Jeffrey and {Chandrasekaran}, Hema and {Twicken}, Joseph D. and {Quintana}, Elisa V. and {Clarke}, Bruce D. and {Allen}, Christopher and {Li}, Jie and {Wu}, Haley and {Tenenbaum}, Peter and {Verner}, Ekaterina and {Bruhweiler}, Frederick and {Barnes}, Jason and {Prsa}, Andrej},
        title = "{Kepler Planet-Detection Mission: Introduction and First Results}",
      journal = {Science},
         year = 2010,
        month = feb,
       volume = {327},
       number = {5968},
        pages = {977},
          doi = {10.1126/science.1185402},
       adsurl = {https://ui.adsabs.harvard.edu/abs/2010Sci...327..977B}
}

@ARTICLE{2015JATIS...1a4003R,
       author = {{Ricker}, George R. and {Winn}, Joshua N. and {Vanderspek}, Roland and {Latham}, David W. and {Bakos}, G{\'a}sp{\'a}r {\'A}. and {Bean}, Jacob L. and {Berta-Thompson}, Zachory K. and {Brown}, Timothy M. and {Buchhave}, Lars and {Butler}, Nathaniel R. and {Butler}, R. Paul and {Chaplin}, William J. and {Charbonneau}, David and {Christensen-Dalsgaard}, J{\o}rgen and {Clampin}, Mark and {Deming}, Drake and {Doty}, John and {De Lee}, Nathan and {Dressing}, Courtney and {Dunham}, Edward W. and {Endl}, Michael and {Fressin}, Francois and {Ge}, Jian and {Henning}, Thomas and {Holman}, Matthew J. and {Howard}, Andrew W. and {Ida}, Shigeru and {Jenkins}, Jon M. and {Jernigan}, Garrett and {Johnson}, John Asher and {Kaltenegger}, Lisa and {Kawai}, Nobuyuki and {Kjeldsen}, Hans and {Laughlin}, Gregory and {Levine}, Alan M. and {Lin}, Douglas and {Lissauer}, Jack J. and {MacQueen}, Phillip and {Marcy}, Geoffrey and {McCullough}, Peter R. and {Morton}, Timothy D. and {Narita}, Norio and {Paegert}, Martin and {Palle}, Enric and {Pepe}, Francesco and {Pepper}, Joshua and {Quirrenbach}, Andreas and {Rinehart}, Stephen A. and {Sasselov}, Dimitar and {Sato}, Bun'ei and {Seager}, Sara and {Sozzetti}, Alessandro and {Stassun}, Keivan G. and {Sullivan}, Peter and {Szentgyorgyi}, Andrew and {Torres}, Guillermo and {Udry}, Stephane and {Villasenor}, Joel},
        title = "{Transiting Exoplanet Survey Satellite (TESS)}",
      journal = {Journal of Astronomical Telescopes, Instruments, and Systems},
         year = 2015,
        month = jan,
       volume = {1},
          eid = {014003},
        pages = {014003},
          doi = {10.1117/1.JATIS.1.1.014003},
       adsurl = {https://ui.adsabs.harvard.edu/abs/2015JATIS...1a4003R}
}

@article{Gaia2016,
  author = {{Gaia Collaboration} and Brown, A. G. A. and Vallenari, A. and et al.},
  title = {The Gaia mission},
  journal = {Astronomy \& Astrophysics},
  year = {2016},
  volume = {595},
  pages = {A2},
  doi = {10.1051/0004-6361/201629272}
}

@ARTICLE{2000A&A...355L..27H,
       author = {{H{\o}g}, E. and {Fabricius}, C. and {Makarov}, V.~V. and {Urban}, S. and {Corbin}, T. and {Wycoff}, G. and {Bastian}, U. and {Schwekendiek}, P. and {Wicenec}, A.},
        title = "{The Tycho-2 catalogue of the 2.5 million brightest stars}",
      journal = {\aap},
         year = 2000,
        month = mar,
       volume = {355},
        pages = {L27-L30},
       adsurl = {https://ui.adsabs.harvard.edu/abs/2000A&A...355L..27H}
}

@ARTICLE{Green2019,
       author = {{Green}, Gregory M. and {Schlafly}, Edward and {Zucker}, Catherine and {Speagle}, Joshua S. and {Finkbeiner}, Douglas},
        title = "{A 3D Dust Map Based on Gaia, Pan-STARRS 1, and 2MASS}",
      journal = {\apj},
         year = 2019,
        month = dec,
       volume = {887},
       number = {1},
          eid = {93},
        pages = {93},
          doi = {10.3847/1538-4357/ab5362},
archivePrefix = {arXiv},
       eprint = {1905.02734},
 primaryClass = {astro-ph.GA},
       adsurl = {https://ui.adsabs.harvard.edu/abs/2019ApJ...887...93G}
}

\end{document}